\documentclass[10pt,twosides,a4paper]{book}
\usepackage{bbm}
\usepackage{bm}
\usepackage{physics} 
\usepackage{amssymb}
\usepackage[pdftex]{graphicx}
\usepackage{latexsym}
\usepackage[small]{caption2}
\usepackage{amsmath}
\usepackage[english]{babel}
\usepackage{babel}
\usepackage{amsmath,amsthm}
\usepackage{rotating}
\usepackage{multirow}
\usepackage{graphicx}
\usepackage{tocloft}
\usepackage{bbm}
\usepackage{dsfont}
\usepackage{amsfonts}
\usepackage{accents}
\usepackage{pdfpages}
\usepackage[thinc]{esdiff}
\usepackage{hyperref}
\usepackage[numbers,sort&compress,square]{natbib}
\usepackage{natbib}
\usepackage{pdfsync}
\usepackage{fancyhdr}
\usepackage{scalerel,stackengine}
\usepackage{fancyhdr}
\usepackage{booktabs}
\renewcommand{\cftchappresnum}{CHAPTER }
\newlength{\mylen}
\fancypagestyle{plain}{} 

\input{tcilatex} 

\begin{document}

\chapter[Fernando C. Lombardo and Paula I. Villar \\
{\em Dynamical Casimir effect in superconducting cavities: from photon creation to universal quantum gates}]{Dynamical Casimir effect in superconducting cavities: from photon generation to universal quantum gates}
\label{chapter1}

\markboth{Dynamical Casimir effect in superconducting cavities}{F. C. Lombardo and P. I. Villar}

\noindent{\large \textbf{Fernando C. Lombardo$^{a}$ and Paula I. Villar$^{b}$}}

\vspace{3mm}

\noindent{ Departamento de F\'\i sica {\it Juan Jos\'e Giambiagi}, FCEyN UBA and IFIBA UBA-CONICET, Facultad de Ciencias Exactas y Naturales, Ciudad Universitaria, Pabell\' on I\smallskip \\ 1428 Buenos Aires, Argentina}



\noindent{\texttt{$^a$lombardo@df.uba.ar; $^{b}$paula@df.uba.ar}}


\section{Introduction}

One of the most remarkable features of quantum field theory is that it predicts the production of particles from the quantum vacuum~\cite{lomb.birrell_davies}. There are notable examples of the conversion of vacuum fluctuations into real particles: Unruh radiation detected by a uniformly accelerated observer~\cite{lomb.unruh}, Hawking radiation originating from a black hole~\cite{lomb.hawking1, lomb.hawking2}, and the Schwinger effect, which produces electron-positron pairs in the presence of a strong electromagnetic field~\cite{lomb.schwinger}. However, while there is strong theoretical support for these effects, none have been experimentally observed. A closely related phenomenon is the dynamical Casimir effect (DCE), which involves the creation of particles from time-dependent external conditions~\cite{lomb.Moore1, lomb.dyncasimir2, lomb.DCE_dalvit, lomb.dodonov_fifty}. It was first proposed in 1970 by Moore, who suggested that in a Fabry-Perot cavity where one of the mirrors oscillates harmonically at twice the frequency of a cavity mode, photons could be generated from the vacuum. Later, it was shown that even a single mirror in free space subjected to non-uniform acceleration would also produce photon radiation~\cite{lomb.FullingDavies}. However, in all cases, the velocity of the mirror required to produce photons was comparable to the speed of light, making it inaccessible for massive moving mirrors~\cite{lomb.lambrecht, lomb.braggio_2005}. More recently, a proposal emerged suggesting that the DCE could be simulated by modifying a field boundary condition~\cite{lomb.dce_nori_pra,lomb.dce_nori_prl}. This idea led to the experimental observation of the DCE. In this setup, the cavity is replaced by a superconducting waveguide terminated by a superconducting quantum interference device (SQUID), and the boundary conditions are modified by applying a time-dependent magnetic flux~\cite{lomb.dce_observacion}. In this way, the DCE has become an indispensable tool for the theoretical and experimental study of the consequences of relativistic dynamics on frontier areas such as quantum thermodynamics~\cite{lomb.DCE_termo1,lomb.DCE_termo2,lomb.DCE_termo3} or relativistic quantum information (RQI).

Although we have stated that the DCE has been experimentally verified in a superconducting circuit, we must point out that this observation lacks a fundamental part of the physics of the effect: the conversion of mechanical energy into photons. Due to this, the experimental realization has sometimes been called a simulation of the effect, and a true observation is still awaited in an optomechanical cavity. Optomechanical systems are composed of an optical cavity formed by two mirrors, one of which is free to move~\cite{lomb.aspelmeyer_review}. Practical photomechanical structures have been obtained, where the mirror oscillates as fast as 6 billion times per second. However, this may not be fast enough; previous studies have shown that the mechanical oscillation frequency must be at least twice the frequency of the cavity's lowest energy mode to observe the DCE. In a recent work~\cite{lomb.prx}, the authors treated both the field inside the cavity and the moving mirror as a quantum system and noted the existence of Casimir-Rabi splitting for mirror frequencies below twice the fundamental mode. The analysis conducted suggested that current optomechanical systems could be used to observe the conversion of mechanical energy into light, meaning that light emission could be achieved in such structures at lower frequencies. Moreover, recent developments in nanoresonator technology~\cite{lomb.nanoresonator}, along with higher quality factor cavities, make the observation of this effect possible in the near future. Another important feature suggested is the fact that the DCE could be analyzed in a more fundamental way with a time-independent Hamiltonian, where an initial state of phonons in the wall evolves into photons inside the cavity. While some extensions of this work have been considered in~\cite{lomb.dce_macri_conversion,lomb.Qin2019}, in this chapter, we do not delve into the quantum transfer of energy from a mechanical degree of freedom to an electromagnetic one (see for instance~\cite{lomb.nos_conversion}). Other proposals considered for observing the conversion of mechanical energy into photons produced by the DCE include hybrid systems of mechanical resonators and superconducting waveguides~\cite{lomb.Motazedifard2018, lomb.Sanz2018}.

The DCE also has interesting connections with quantum thermodynamics. This area of study in physics arises from the relentless miniaturization of technological devices~\cite{lomb.ion,lomb.circuitqed_review} and encapsulates two different yet complementary aspects. On the one hand, it aims to derive the laws of thermodynamics rigorously from microscopic quantum interactions. On the other hand, from a more applied perspective, it seeks to enhance thermodynamic processes, such as the conversion of heat into work, using quantum phenomena with no classical analogue, such as coherence~\cite{lomb.coherence1,lomb.coherence2} or entanglement~\cite{lomb.entanglement1,lomb.entanglement2}. The concept of information, and its intimate connection with entropy and thermodynamics, plays a crucial role in both aspects~\cite{lomb.information}.

Most research in this area has been implemented on qubits~\cite{lomb.ottoqubit} or harmonic oscillators subjected to different thermodynamic cycles~\cite{lomb.otto_kosloff}. While in certain cases a quantum field in a cavity can be studied as a few modes that behave like harmonic oscillators, there are significant circumstances under which this approximation fails. However, only a handful of works have studied the effects arising from a full quantum field~\cite{lomb.qftmachine,lomb.unruhengine1,lomb.unruhengine2,lomb.unruhengine3}, and most of them treat it as a bath rather than a working medium. For this reason, in this chapter, we will investigate the operation of a thermal machine based on a thermodynamic cycle of a quantum field.

One of the difficulties that arises with finite-time operation of these quantum thermal machines is the emergence of coherences in the system's state, resulting in a loss of efficiency~\cite{lomb.ottoqubit, lomb.otto_kosloff, lomb.Otto_nos}. In response to this problem, protocols called “shortcut-to-adiabaticity” (STA) have emerged, allowing the system to reach the final state that would have been obtained through an infinitely slow, adiabatic evolution, but in a finite time~\cite{lomb.STA_review}. In general, this means that no new excitations will be generated in the final state; however, it is worth noting that some STA methods, such as “transitionless quantum driving”~\cite{lomb.berry, lomb.delcampo_prl}, also ensure the suppression of non-adiabatic excitations during intermediate times. Other methods for implementing STA include the use of fast-forward protocols~\cite{lomb.FF}, invariants~\cite{lomb.muga_prl}, optimal protocols~\cite{lomb.OCT}, fast quasi-adiabatic protocols, etc. A disadvantage of these methods is that they typically require full quantum control of the system, making their implementation extremely challenging from an experimental perspective.

STAs have been considered both theoretically and experimentally for a variety of physical systems, such as trapped ions~\cite{lomb.fast_transport_ion}, cold atoms~\cite{lomb.fast_expansion_atom}, ultra-cold Fermi gases~\cite{lomb.fast_control_gas}, Bose-Einstein condensates on atom chips~\cite{lomb.control_einstein_condensate}, spin systems~\cite{lomb.spin_engine_STA}, and others. In Ref.~\cite{lomb.nos_STA1} we have described how to implement STA for the case of a massless scalar field inside a cavity with a moving wall, in $1+1$ dimensions. The approach is based on the known solution to the problem that exploits the conformal symmetry, and the shortcuts take place whenever there is no DCE. We have also described possible experimental realizations of the shortcuts using superconducting circuits (see~\cite{lomb.nos_STA2,lomb.nos_STA3}).

Regarding thermal machines, these protocols have been proposed to alleviate the trade-off between efficiency and power~\cite{lomb.STA_otto,lomb.STA_otto2,lomb.STA_otto_cost}, both for quantum engines of a single particle~\cite{lomb.delcampo_more_bang} and for many-particle systems~\cite{lomb.delcampo_many_particle1,lomb.delcampo_many_particle2,lomb.Fogarty1, lomb.Fogarty2}. An experiment has even been conducted with a Fermi gas implementing these protocols~\cite{lomb.delcampo_experiment}. Additionally, STA have been derived for relativistic quantum systems evolving under Dirac dynamics~\cite{lomb.deffner_dirac1,lomb.deffner_dirac2}. Despite their effectiveness in improving the efficiency of thermal machines, shortcuts to adiabaticity also incur an energy cost that must be considered~\cite{lomb.STA_cost, lomb.STA_calzetta}, which can be used to establish a connection between quantum engines and the third law of thermodynamics~\cite{lomb.third_law_muga}.

Another cutting-edge field where the DCE plays a significant role is circuit quantum electrodynamics (cQED), where the effect was observed for the first time. This area has recently become the leading architecture for quantum computing~\cite{lomb.Blais2020}. In such systems, one can control the interaction between artificial atoms (or qubits in the two-level approximation) and the electromagnetic field~\cite{lomb.circuitqed_review}. Photons are stored in a microwave resonator, and the qubits are designed using circuits that include Josephson junctions, capacitors, and/or inductors.

This architecture has flourished recently with the emergence of quantum engineering technologies, not only due to its success in manipulation and measurement but also because of its scalability properties. In fact, cQED is one of the main candidates for achieving scalable quantum computing and has been used to manipulate dozens of qubits, performing tasks that bring this technology close to the goal of quantum supremacy~\cite{lomb.Arute2019}. It has also been used to test key ideas in quantum error correction and the implementation of fault-tolerant architectures~\cite{lomb.error1,lomb.error2,lomb.error3,lomb.error4,lomb.error5,lomb.error6,lomb.error7}. Later, we will show how to use this architecture to implement quantum gates and generate states that are robust against errors.

In summary, the structure of this chapter is as follows. In section~\ref{lomb.chap:DEC}, we review the origin and fundamental properties of the DCE, as well as three equivalent mathematical descriptions of the phenomenon that will be employed later. Section~\ref{lomb.chap:DCEcQED} is dedicated to the implementation of the DCE in architectures based on cQED. In section~\ref{lomb.chap:otto}, we will investigate the impact of the DCE on the performance of a thermal machine based on a quantum field. Along section~\ref{lomb.chap:ctlsqz}, we will see how the DCE can be used to implement a controlled-squeeze gate in a cQED architecture. Finally, in~\ref{lomb.chap:summary}, we will present a summary.

\section{Dynamical Casimir effect}
\label{lomb.chap:DEC}
\subsection{Cavity with moving mirrors}

\begin{figure}[!htb]
\begin{center}
\includegraphics[width=0.6\textwidth]{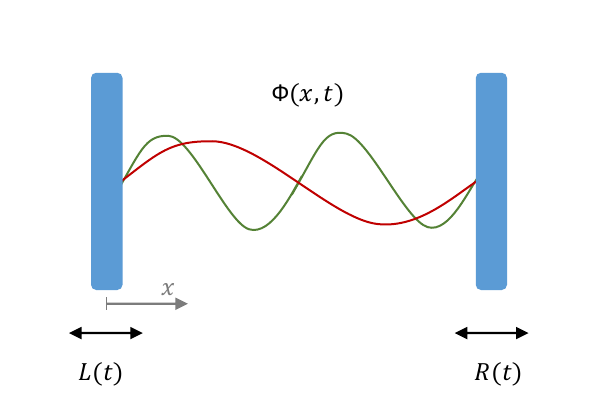} {}
\end{center}
\caption{Diagram of a cavity with moving mirrors following the trajectories $L(t)$ and $R(t)$.}
\label{lomb.cavidad}
\end{figure}

The DCE consists of particle generation from the vacuum due to the temporal variation of boundary conditions applied to a quantum field. Let us start with a model that includes a massless scalar field $\Phi(x,t)$ within a one-dimensional cavity formed by two perfectly reflective, flat, parallel, infinite mirrors located at $x=L(t)$ and $x=R(t)$, which move according to prescribed trajectories (Fig.~\ref{lomb.cavidad}). Analogous to the electromagnetic field, we assume that its dynamics follow the wave equation.
\begin{equation}
    \partial_x^2 \Phi(x,t)- \partial_t^2 \Phi(x,t)=0
\end{equation}
with Dirichlet boundary conditions on each mirror
\begin{equation}
    \Phi(L(t),t)=\Phi(R(t),t)=0.
\end{equation}
In the previous equation and throughout this  chapter, we will use natural units where $\hbar = c = k_B = 1$, unless otherwise specified. The justification for this model lies in the fact that the transverse-electric (TE) and transverse-magnetic (TM) modes of the electromagnetic field can be described by scalar fields~\cite{lomb.crocce_2001}.

The quantization of fields with time-dependent boundary conditions, as with curved spaces, tends to be delicate~\cite{lomb.birrell_davies}. This is because there is no exact Hamiltonian formulation of the problem; if such a formulation existed, a unitary time evolution operator would exist such that
\begin{equation}
    \hat{\Phi}(x,t)=U^\dagger(t,t_0)\hat{\Phi}(x,t_0)U(t,t_0),
\end{equation}
but if the mirror was at \((x, t_0)\), then we would have \(\hat{\Phi}(x, t) = 0\), which is absurd. We therefore conclude that such a unitary operator does not exist and, consequently, neither does a Hamiltonian that generates the time evolution; thus, there is no Schrödinger picture.

However, for early times, before the mirrors begin to move, the cavity remains static, and the field can be expanded as
\begin{equation}
\hat{\Phi}(x,t)=\sum_{n=0}^\infty\left[\hat{a}_{n}^{\text{in}}u_{ n}({ x},t)+h.c.\right],
\label{lomb.chap:repaso-eq:modos_in}
\end{equation}
Where \(\hat{a}_{n}^{\text{in}}\) are the bosonic operators corresponding to photons of different frequencies, and the functions \(u_{n}\) are the positive-frequency solutions of the wave equation:

\begin{equation}
u_{n}({x},t) = \sqrt{\frac{2}{d(0)}} \sin(k_n x) \frac{e^{-i\omega_{n} t}}{\sqrt{2\omega_{n}}}, \label{lomb.chap:repaso-eq:u_n}
\end{equation}
where \({k}_n = \frac{n\pi}{d(0)}\) is the wavenumber, \(\omega_{n} = k_n\) is the frequency, and \(d(t) = R(t) - L(t)\) is the cavity length. The mirrors start to move at \(t = 0\), and this initial basis is continuously deformed, through the equation of motion, into another that satisfies the new boundary conditions. Once stopped in their original positions, quantization and particle definitions are again well-defined, but now we have a new basis \(v_{k}({x},t)\) and new final bosonic operators \(\hat{a}_{n}^{\text{out}}\), so the field will expand as
\begin{equation}
\hat{\Phi}({x},t) = \sum_{n=0}^\infty \left[ \hat{a}_{n}^{\text{out}} v_{n}({x},t) + h.c. \right]. \label{lomb.chap:repaso-eq:modos_out}
\end{equation}

However, the field can be alternatively expressed in this expansion, resulting from time evolution, as well as in the initial basis. Thus, there exist coefficients \(\alpha_{nk}\) and \(\beta_{nk}\) such that
\begin{equation}
v_{n} = \sum_{k=0}^\infty \left[\alpha_{nk} u_{k} + \beta_{nk} u_{k}^* \right]. \label{lomb.chap:repaso-eq:vn_un}
\end{equation}

Using this,  we can relate both sets of bosonic operators as
\[
\hat{a}_{n}^{\text{out}} = \sum_{k} \left[\alpha_{nk} \hat{a}_{k}^{\text{in}} + \beta_{nk}^* \hat{a}_{k}^{\text{in}\dagger}\right],
\]
which is known as a Bogoliubov transformation. It is important to note that these are Gaussian transformations, meaning they map Gaussian states (such as coherent, thermal, and squeezed states) to other Gaussian states.

Later on, we will see different methods to calculate the Bogoliubov coefficients, but we can already appreciate the fundamental property of the DCE. To do so, let us consider an initial state given by the density matrix \(\hat{\rho}\) and calculate the mean photon number in mode \(k\) at the end of the motion, \(N_k = \text{Tr}(\hat{\rho} \hat{N}_k)\), expressing the final operators in terms of the initial ones:
\[
N_k^{\text{out}} = \sum_{{n}=0}^\infty \left(|\alpha_{{nk}}|^2 N_{{n}}^{\text{in}} + |\beta_{{nk}}|^2 N_{{n}}^{\text{in}} + |\beta_{{nk}}|^2 \right),
\]
where \(N_n^{\text{in}}\) is the initial mean photon number in mode \(n\). We then observe that for \( |\beta_{nk}| \neq 0 \), photons are generated in mode \({n}\) even if the initial photon number is zero, i.e., from the vacuum.

\subsection{Bogoliubov coeffients}

In the following, we will review the calculation of the Bogoliubov coefficients, which will allow us to understand more deeply the physical origin of the DCE. We will do this in the case where only the right wall is moving, taking \(L(t)=0\). To obtain these coefficients, it is necessary to evolve the initial solution basis \(u_n\) in time to the final basis \(v_n\) and then rewrite the latter in terms of the former.

To do this, we can express the new basis of solutions in the instantaneous basis given by
\begin{equation}
\label{lomb.chap:repaso-eq:base_inst}
    \psi_{j}^{\prime\prime}(x)+k_{j}\psi_{j}(x)=0
\end{equation}
and Dirichlet boundary conditions
\begin{equation}
    \psi_{j}(0)=\psi_{j}(R(t))=0,
\end{equation}
where \(k_j(R)=\frac{j\pi}{R}=\omega_j(R)\). If we further ask it to be an orthonormal basis at each instant with respect to the inner product
\begin{equation}
\label{lomb.chap:repaso-pi1D}
\langle\psi_k,\psi_j\rangle=\int_{0}^{R(t)}\psi_k(x,R(t))\psi_j(x,R(t))dx=\delta_{kj},
\end{equation}
then the functions take the following form
\begin{equation}
\psi_{ j}({ x},R)=\sqrt{\frac{2}{R}}\sin\left[k_j(R)x\right].
\end{equation}

This allows us to expand the final basis as
\begin{equation}
\label{lomb.chap:repaso-eq:Qkn}
v_{ n}({ x},t)=\sum_{ k}\left[Q_{ k}^{({ n})}(t)\psi_{ k}(x,R(t))+Q_{ k}^{({ n})*}(t)\psi_{ k}(x,R(t))\right]
\end{equation}
where \(Q_{ k}^{({ n})}(t)\) are time-dependent coefficients. These can be found by substituting Eq.~(\ref{lomb.chap:repaso-eq:Qkn}) into Eq.~(\ref{lomb.chap:repaso-eq:modos_out}) and then substituting the field into the wave equation. Next, using the orthogonality of the \(\psi_j\), we can provide a set of coupled differential equations
\begin{equation}
\label{lomb.chap:repaso-eq:dif_Qkn}
\ddot{Q}_{k}^{(n)}+\omega_{k}^{2}(t){Q}_{k}^{(n)}=2\dot{R}\sum_{j}M_{kj}\dot{Q}_{j}^{(n)}+\ddot{R}\sum_{j}M_{kj}{Q}_{j}^{(n)}+\dot{R}^{2}\sum_{j}S_{kj}{Q}_{j}^{(n)},
\end{equation}
where we define the coupling coefficients
\begin{align}
    M_{kj}&=\langle\psi_{j},\partial_{R}\psi_{k}\rangle =-M_{jk}\\
    S_{kj}&=\langle\partial_{R}\psi_{j},\partial_{R}\psi_{k}\rangle=\sum_lM_{lk}M_{lj},
\end{align}
noting that the second equality for \(M_{kj}\) comes from writing the inner product as an integral and integrating by parts, while the second equality comes from the completeness of the solution basis. These coefficients can be explicitly calculated using the expression for \(\psi_j\). Note that, for both early and late times, the cavity is static, and thus the right-hand side of Eq.(\ref{lomb.chap:repaso-eq:dif_Qkn}) vanishes, resulting in a solution that is that of a harmonic oscillator with frequency \(\omega_k\). Then, if we set the initial conditions
\begin{align}
\label{lomb.chap:repaso-eq_Qkn_inicial}
    Q_k^{(n)}(0)&=\frac{1}{\sqrt{2\omega_k}}\delta_{nk}\\
    \dot{Q}_k^{(n)}(0)&=\frac{-i}{\sqrt{2\omega_k}}\delta_{nk}
\end{align}
we will find that for early times
\begin{equation}
    Q_k^{(n)}(t<0)=\frac{e^{-i\omega_kt}}{2\omega_k},
\end{equation}
which means that this expansion coincides with the initial basis $Q_k^{(n)}(t)\psi_k(x,R(t))= u_k(x,t)$; while for late times we will again have a harmonic oscillator
\begin{equation}
\label{lomb.chap:repaso-eq:Q_kn_tf}
    Q_k^{(n)}(t>t_f)=\frac{1}{2\omega_k}\left(\alpha_{nk}e^{-i\omega_kt}+\beta_{nk}e^{i\omega_kt}\right),
\end{equation}
where the coefficients \(\alpha_{nk}\) and \(\beta_{nk}\) are obtained by solving the differential equation for the \(Q_k^{(n)}\). Furthermore, if we substitute Eq.~(\ref{lomb.chap:repaso-eq:Q_kn_tf}) into Eq.~(\ref{lomb.chap:repaso-eq:Qkn}) and compare with Eq. (\ref{lomb.chap:repaso-eq:vn_un}), it is clear that these coefficients are precisely those of Bogoliubov.

We have thus reduced the original problem of a quantum scalar field in a cavity with a moving wall to a system of second-order ordinary differential equations. This can be done numerically for any trajectory of the mirror by truncating the number of modes, but there are also analytical methods, such as multiple scale analysis (MSA), which we will analyze next, that allow for an approximate solution for certain trajectories of interest.

\subsection{Beyond perturbative treatment: Multiple scale analysis}

It is important to highlight that, while from a theoretical point of view, as long as the Bogoliubov coefficient $\beta_{nk}$ does not vanish, there will be vacuum photon creation, there are certain trajectories that allow for a significant increase in this effect. Below, we will see that this is the case for harmonic oscillatory trajectories of the mirror of the form \begin{equation} R(t)=R_0(1+\epsilon\sin \Omega t), \end{equation} where $R_0$ is the initial and equilibrium position, $\epsilon\ll1$ is a small dimensionless parameter that determines the amplitude of the oscillation, and $\Omega$ is the oscillation frequency.

At this point, one possibility for solving Eq.~(\ref{lomb.chap:repaso-eq}) would be to perform a perturbative expansion in $\epsilon$ and solve order by order in this variable. However, this method is only valid for short times such that $\Omega\epsilon t\ll1$, since ``secular'' terms appear in the equations that, at long times, cause one order to have a magnitude comparable to that of the next, breaking the approximation. To extend to longer times, we can use the method of multiple scales, which consists of defining a slower time scale $\tau= \epsilon t$ and making the following expansion
\begin{equation} \label{lomb.chap:repaso-eq} Q_k^{(n)}(t,\tau)=\frac{1}{2\omega_k}\left(\alpha_{nk}(\tau)e^{-i\omega_kt}+\beta_{nk}(\tau)e^{i\omega_kt}\right),
\end{equation}
Then, if we expand Eq.~(\ref{lomb.chap:repaso-eq:dif_Qkn}) to first order in $\epsilon$, multiply by $e^{-i\omega_kt}$ and $e^{i\omega_kt}$, and integrate, we obtain linear differential equations for the Bogoliubov coefficients.
\begin{align} \label{lomb.chap:repaso-eq2}
\beta_{nk}^{\prime}=&\frac{\omega_{k}^{\prime}R_{0}}{2}\alpha_{nk}\delta(\Omega-2\omega_{k})+\frac{R_{0}\Omega}{2\omega_{k}}\sum_{j}M_{kj}\left(\left(-\omega_{k}+\frac{\Omega}{2}\right)\alpha_{nj}\delta(-\omega_{j}-\omega_{k}+\Omega)\right)\nonumber\\ +&\frac{R_{0}\Omega}{2\omega_{k}}\sum_{j}M_{kj}\left(\left(\omega_{j}+\frac{\Omega}{2}\right)\beta_{nj}\delta(\omega_{j}-\omega_{i}+\Omega)+\left(\omega_{j}-\frac{\Omega}{2}\right)\beta_{nj}\delta(\omega_{j}-\omega_{k}-\Omega)\right)\\
\alpha_{nk}^{\prime}=&\frac{\omega_{k}^{\prime}R_{0}}{2}\beta_{nk}\delta(\Omega-2\omega_{k})+\frac{R_{0}\Omega}{2\omega_{k}}\sum_{j}M_{kj}\left(\left(-\omega_{k}+\frac{\Omega}{2}\right)\beta_{nj}\delta(-\omega_{j}-\omega_{k}+\Omega)\right)\nonumber\\
+&\frac{R_{0}\Omega}{2\omega_{k}}\sum_{j}M_{kj}\left(\left(\omega_{j}+\frac{\Omega}{2}\right)\alpha_{nj}\delta(\omega_{j}-\omega_{i}+\Omega)+\left(\omega_{j}-\frac{\Omega}{2}\right)\alpha_{nj}\delta(\omega_{j}-\omega_{k}-\Omega)\right)
\end{align}
where we take $\delta(x)=1$. Let us also remember that the initial conditions for $Q_k^{(n)}$, Eq.~(\ref{lomb.chap:repaso-eq_Qkn_inicial}), imply the following initial conditions for the Bogoliubov coefficients
\begin{align} \label{lomb.chap}
\alpha_{nk}(0)&=\delta_{nk}\\
\beta_{nk}(0)&=0. \end{align}

From these equations, even without solving them, we can already draw several interesting physical conclusions. On one hand, since the $\beta_{nk}$ are initially zero and their derivative depends on the $\alpha_{nk}$ through two delta terms, they will not vanish only for:
\begin{align} \Omega&=2\omega_k\nonumber\\ \Omega&=\omega_k+\omega_j.
\end{align}
In other words, under this approximation, photon creation will only occur if these conditions of {\bf parametric resonance} are satisfied. Meanwhile, if $\Omega=|\omega_k-\omega_j|$, only the coefficients $\alpha_{nk}$ will evolve over time, leading to a redistribution of the initial photons from one mode to others \cite{lomb.nos_entangled}. We can also note that the cavity spectrum becomes fundamental to understanding photon generation, as it will determine which and how many terms contribute to the differential equation of the Bogoliubov coefficients.

\subsection{Hamiltonian formalism}

There are many cases where the previous formulation in terms of the Bogoliubov coefficients is sufficient to provide a complete and accurate description of the system. However, it only describes the evolution of the creation and annihilation operators, while for a quantum treatment, it would be useful to understand the dynamics of the states in Hilbert space. The problem is that, as mentioned earlier, there is no unitary operator that exactly describes this system. Still, we will see that it is possible to construct a Hamiltonian that generates an approximate unitary operator that describes the dynamics of the states correctly (as it is done in~\cite{lomb.nos_conversion}).

To do this, we can start by writing the Lagrangian that gives rise to the wave equation \begin{equation} \label{lomb.chap:repaso-eq
} L=\frac{1}{2}\int_{0}^{R(t)}(\partial_t^2\Phi(x,t)-\partial_x^2\Phi(x,t))dx. \end{equation} If we now expand the field again in an instantaneous basis \begin{equation} \Phi(x,t)=\sum_{n=1}^{\infty}\left[Q_{n}(t)\psi_{n}(x,R(t))+ \text{h.c.} \right], \end{equation} where the functions $\psi_n(x,R(t))$ are the same as in Eq.~(\ref{lomb.chap:repaso-eq:base_inst}), and using the properties from the previous section, we have \begin{equation} L=\frac{1}{2}\sum_{k}\left(\dot{Q_{k}}^{2}-\omega_{k}^{2}Q_{k}^{2}\right)-\dot{R}\sum_{k,j}M_{kj}\dot{Q_{k}}Q_{j}+\frac{\dot{R}^{2}}{2}\sum_{k,j,l}M_{jk}M_{lk}Q_{k}Q_{j}. \end{equation} It is worth noting that this Lagrangian gives rise to motion equations for the coefficients $Q_k$ identical to Eq. (\ref{lomb.chap:repaso-eq
}). Next, we calculate the conjugate canonical momentum \begin{equation} P_{k}=\frac{\partial L}{\partial\dot{Q}{k}}=\dot{Q{k}}-\dot{R}\sum_{l,j}M_{lj}\delta_{lk}Q_{j}=\dot{Q_{k}}-\dot{R}\sum_{j}M_{kj}Q_{j} \end{equation} which we can substitute to obtain the classical Hamiltonian of the system \begin{equation} H=\sum_{i}P_{i}\dot{Q}{k}-L\ =\frac{1}{2}\sum{k}\left[P_{k}^{2}+\omega_{k}^{2}Q_{k}^{2}\right]+\dot{R}\sum_{k,j}M_{kj}P_{k}Q_{j}. \end{equation} The quantum Hamiltonian can be obtained through canonical quantization, promoting the conjugate canonical variables to operators and establishing the usual commutation rules \begin{equation} Q_{k}\to \hat{Q}_k, \quad P_{k}\to \hat{P}_k,   \quad [\hat{Q}_k,\hat{P}_{j}]=i\delta_{kj}. \end{equation} Now we can define the creation $\hat{a}_k^\dagger$ and annihilation $\hat{a}_k$ operators as \begin{align} \hat{a}_{k}=\sqrt{\frac{1}{2\hbar\omega_{k}(R)}}(\omega_{k}(R\hat{Q}_{k}+i\hat{P}_{k}), \end{align} and from them, reproduce the usual Hilbert space defined as $|n\rangle=(a^{\dagger})^n|0\rangle/\sqrt{n!}$.

We can also see the equivalence with the previous approach by expressing the final operators (in the Heisenberg picture) in terms of the initial ones \begin{equation}
    \hat{Q}_{n}(t)=\sum_{k}\left[Q_{n}^{(k)}(t)\hat{a}_{k}+Q_{n}^{(k)*}(t)\hat{a}_{k}^{\dagger}\right]
\end{equation} where the functions $Q_{n}^{(k)}(t)$ must be a basis of solutions of Eq. (\ref{lomb.chap:repaso-eq:dif_Qkn}) such that
\begin{equation}
    \hat{Q}_{n}(0)=\sum_{k}\left[Q_{n}^{(k)}(0)\hat{a}_{k}+Q_{n}^{(k)*}(0)\hat{a}_{k}^{\dagger}\right]=\frac{1}{\sqrt{2\omega_{n}}}\left(\hat{a}_{n}+\hat{a}_{k}^{\dagger}\right)
\end{equation}
\begin{equation}
    \hat{P}_{k}(0)=\dot{\hat{Q}}_{n}(0)=\sum_{k}\left[\dot{Q}_{n}^{(k)}(0)\hat{a}_{k}+\dot{Q}_{n}^{(k)*}(0)\hat{a}_{k}^{\dagger}\right]=\frac{-i}{\sqrt{2\omega_{n}}}\left(\hat{a}_{n}-\hat{a}_{k}^{\dagger}\right),
\end{equation}
implying that they must satisfy initial conditions in order to match the functions from the previous section and, except for a factor, with the Bogoliubov coefficients. Thus, it is clear that the subscript is already present in the classical dynamics while the superscript arises from the necessity in quantum mechanics to solve the classical equations for all possible initial conditions.

Having understood the equivalence with the previous section, we can write the Hamiltonian in terms of the creation and annihilation operators as
\begin{equation} H=\sum_{k}\omega_{k}a_{k}^{\dagger}a_{k}+\dot{R}\sum_{j,k}M_{jk}\frac{1}{2i}\sqrt{\frac{\omega_{j}}{\omega_{k}}}\left(a_{j}-a_{j}^{\dagger}\right)\left(a_{k}+a_{k}^{\dagger}\right). \end{equation}
If we also consider a trajectory with a small variation around its equilibrium position
\begin{equation} R(t)\approx R_{0}+\delta R(t) \end{equation}
and expand the creation operators to first order \begin{equation} \label{lomb.chap:primerorden} a_{k}(R)\approx a_{k}+\delta R\frac{\omega_{k}^{\prime}}{2\omega_{k}}a_{k}^{\dagger},
\end{equation}
we obtain the Hamiltonian
\begin{align} \label{lomb.chap:repaso-eq}
H=&\sum_{k}\omega_{k}a_{k}^{\dagger}a_{k}+\sum_{k}\omega_{k}^{\prime}\delta Ra_{k}^{\dagger}a_{k}+\sum_{k}\frac{\omega_{k}^{\prime}}{2}\delta R(a_{k}^2+a_{k}^{\dagger 2})\nonumber\\
&+\dot{R}\sum_{j,k}M_{jk}\frac{1}{2i}\sqrt{\frac{\omega_{j}}{\omega_{k}}}\left(a_{j}a_{k}-a_{j}^{\dagger}a_{k}+a_{j}a_{k}^{\dagger}-a_{j}^{\dagger}a_{k}^{\dagger}\right)+\mathcal{O}(\delta R^{2}). \end{align}
In this Hamiltonian, we can identify three contributions. On one hand, the first term of the Hamiltonian is the free energy of the harmonic oscillator at the reference position, while the second is a correction due to the fact that the wall movement modifies the frequency of the photons. The third is the essence of the DCE and evidences the creation of photon pairs, or a one-mode squeeze, in one mode of the field, while the last term contains the generation of photon pairs in two modes $a_k^\dagger a_j^\dagger$, or a two-mode squeeze operation (both associated with the $\beta_{nk}$ of the previous section). Finally, we have terms $a_k a_j^\dagger$ that allow the scattering of photons from one mode to another while preserving the total number (associated with the $\alpha_{nk}$ of the previous section).

This Hamiltonian method and the Bogoliubov coefficients method are easily generalizable to a 3D cavity, or to a massive scalar field, or to more general boundary conditions; both for an optomechanical cavity and for superconducting circuits.

\subsection{Moore's functions}

The two methods discussed so far allowed us to provide a description of the DCE in terms of the temporal evolution of Bogoliubov coefficients and, consequently, of the creation and annihilation operators for each mode. However, these descriptions have their limitations, as studying phenomena localized in space becomes extremely difficult at both the theoretical and computational levels when the modes are delocalized throughout the cavity. For instance, if we consider only \(N\) modes, we must solve that number of differential equations. Fortunately, in the case of a massless scalar field with a moving boundary condition in one dimension, there exists an alternative description via Moore functions.

Under these hypotheses, the basic observation that allows us to provide a new mathematical description of the system is conformal symmetry. Recall that this means that if we make the change of variables
\begin{equation}\label{lomb.chap:repaso-eq:conftransf}
\bar t +\bar x =G(t+x),\quad \quad \bar t -\bar x = F(t-x),
\end{equation}
where \(F\) and \(G\) are functions whose derivatives never vanish, then the field in the new variables \(\bar{\Phi}(\bar{x},\bar{t})=\Phi(x,t)\) continues to satisfy the wave equation
\begin{equation}
    (\partial_{\bar x}^2-\partial_{\bar t}^2)\bar{\Phi}(\bar{x},\bar{t})=0.
\end{equation}

For our problem, these functions provide us with two degrees of freedom that allow us to map two points to two arbitrary points. Thus, we can choose Moore functions that at each instant map the edges of the cavity to the same points; for example, the left edge at \(L(t)\) to \(\bar{x}=0\) and the right edge to \(\bar{x}=1\), which reduces the problem in the new coordinates to solving the wave equation in a static cavity. Mathematically, we can find these transformations by subtracting Eqs.~(\ref{lomb.chap:repaso-eq:conftransf}) and substituting, yielding the Moore equations
\begin{align}
\label{lomb.chap:repaso-eq:MooreEqL}
    G(t+L(t))-F(t-L(t))&=0\\
G(t+R(t))-F(t-R(t))&=2. \label{lomb.chap:repaso-eq:MooreEqR}
\end{align}
As we mentioned, in these conformal coordinates, the solution of the field can be easily found, and returning to the real coordinates, the field solution can be expressed as
\begin{equation}
\label{lomb.chap:repaso-eq:field_Moore}
    \Phi(x,t)=\sum_{n=1}^{\infty}\left[a_n v_n(x,t)+a_n^\dagger v_n^*(x,t)\right],
\end{equation}
where the modes are given by \cite{lomb.dalvit_two_mirrors}
\begin{equation}
    v_n(x,t)=\frac{i}{\sqrt{4\pi n}}[e^{-in\pi G(t+x)}+e^{in\pi F(t-x)}].
\end{equation}

If we consider, as in previous sections, that \(L(t)=0\), we have \(F(z)=G(z)\), and we obtain only one Moore equation
\begin{align}
F(t+R(t))-F(t-R(t))=2. \label{lomb.chap:repaso-eq:Moore_R0}
\end{align}

Then, to understand how photon generation is observed in this formalism, it is useful to solve the Moore equation for early and late times. In both cases, the position of the wall is constant, \(R\neq R(t)\), and we must solve the equation
\begin{align}
F(t+R)-F(t-R)=2.
\end{align}
The solution has the form
\begin{equation}
    F(z)=\frac{z}{R}+g(z),
\end{equation}
i.e., a linear function plus another \(R\)-periodic \(g(z)\). For early times, \(t<0\), this function can be determined by requiring that the previous basis \(v_n\) coincides with the \(u_n\) from Eq.~(\ref{lomb.chap:repaso-eq:u_n}), from which we obtain
\begin{equation}
    F(z<0)=\frac{z}{R(0)},
\end{equation}
thus nullifying the periodic function. Then, taking this as an initial condition and solving the Moore equation from beginning to end, we will determine the periodic function \(g\) for late times. In general, it will be non-zero, giving rise to a new basis and, consequently, Bogoliubov coefficients \(\beta_{nk}\neq0\) that signal the creation of photons due to the DEC.

Additionally, using the expansion of Eq.~(\ref{lomb.chap:repaso-eq:field_Moore}), we can calculate the renormalized energy-momentum tensor \cite{lomb.dalvit_two_mirrors}. This can be done using the standard approach based on point-splitting regularization (see for example~\cite{lomb.FullingDavies}), but it can also be derived using the conformal anomaly associated with the conformal transformation~\cite{lomb.birrell_davies,lomb.libro_calzetta}. In any case, we obtain the energy density of the field
\begin{align}
\label{lomb.chap:repaso-eq:dens_en}
\langle T_{tt}(x,t)\rangle_{\text{ren}} = f_G(t+x)+f_F(t-x),
\end{align}
where
\begin{eqnarray}
f_G &=& -\frac{1}{24\pi}\left[\frac{G'''}{G'}-\dfrac{3}{2}\left(\frac{G''}{G'}\right)^2\right]+\dfrac{(G')^2}{2}\left[-\dfrac{\pi}{24}+Z(Td_0)\right] \nonumber\\
f_F &=& -\frac{1}{24\pi}\left[\frac{F'''}{F'}-\dfrac{3}{2}\left(\frac{F''}{F'}\right)^2\right]+\dfrac{(F')^2}{2}\left[-\dfrac{\pi}{24}+Z(Td_0)\right]    ,
\end{eqnarray}
with \(d_0=|R(0)-L(0)|\) being the initial length of the cavity. We are considering that the initial state of the field is thermal at temperature \(T\), and the factor \(Z(Td_0)\) is related to the initial average energy
\begin{equation}
Z(Td_0)=\sum_{n=1 }^\infty\frac{n\pi}{\exp\left(\frac{n\pi}{Td_0}\right)-1}.
\end{equation}
These equations are a slight generalization of the results for zero temperature from~\cite{lomb.FullingDavies}, see also~\cite{lomb.DCE_temperatura_alves}.
Finally, we can note that when the Moore function is linear, the renormalized energy density reduces to the energy of the static Casimir effect. This again indicates that particle creation occurs when the functions \(F(z)\) and \(G(z)\) are non-linear.

This method is evidently more efficient than either of the two previous ones since it does not involve solving a system of infinite coupled differential equations or calculating the temporal evolution of a time-dependent Hamiltonian, but rather solving at most two equations for the Moore functions. The cost of this enormous simplification is the inability to generalize this method to new systems that do not possess conformal symmetry, such as a three-dimensional cavity, a massive scalar field, or more general boundary conditions.

\section{DEC in superconducting quantum circuits}
\label{lomb.chap:DCEcQED}

From an experimental point of view, the original proposal to observe the DCE in an optomechanical cavity faces serious problems. The most important issue is that the number of photons produced is proportional to the ratio between the velocity of the mirror and the speed of light, meaning that generating a measurable amount of photons requires moving the wall at a large fraction of the speed of light, which is extremely difficult to achieve experimentally for a massive mirror. For this reason, it has been suggested that the DCE essentially consists of the generation of excitations of a quantum field with a time-dependent boundary condition. In the case of optomechanical cavities, the field is the electromagnetic one, and the moving mirror generates the time-dependent boundary condition. Based on this principle, an equivalent system has been proposed, consisting of a superconducting microwave waveguide terminated in a SQUID subjected to a time-dependent magnetic field. The main advantage of this system is that the boundary condition does not require the ultra-fast movement of a massive object, but simply a rapid variation of a magnetic field, which is much more accessible since GHz frequencies are common in such circuits.

\begin{figure}[!htb]
\begin{center}
\includegraphics[width=0.60\textwidth]{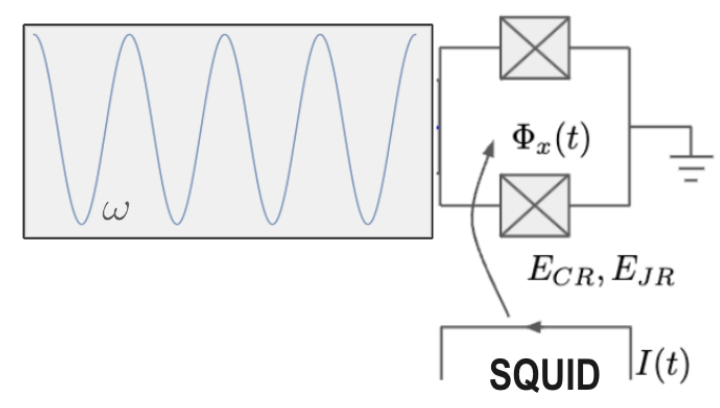} {}
\end{center}
\caption{Diagram of a one-dimensional SQUID-terminated resonator.}
\label{}
\end{figure}

Mathematically, the equivalence can be seen by starting from the Lagrangian of the phase field \(\Phi(x,t)\) in a superconducting waveguide of length \(L_0\) with capacitance \(c\) and inductance \(l\) per unit length, terminated in a SQUID, as given by~\cite{lomb.shumeiko_2013}
\begin{eqnarray}\label{lomb.chap:repaso-eq:lagrangiano-super}
L_{\rm cav} &=& \left(\frac{\hbar}{2e}\right)^2 \frac{c}{2} \int_0^{L_0} d x \left((\partial_t \Phi)^2 - v^2 (\partial_x \Phi)^2 \right)\\
&+& \left[ \left(\frac{\hbar}{2e}\right)^2 \frac{2 C_J}{2} \partial_t \Phi(L_0,t)^2
 -  E_J \cos{f(t)} \Phi(L_0,t)^2
 \right]
 \,\nonumber ,
\end{eqnarray}
where \(v = 1/\sqrt{l c}\) is the propagation speed of the field and \(f(t)=\frac{2e }{\hbar}\phi_x(t)\) is the phase across the SQUID controlled by the external magnetic flux \(\phi_x(t)\). Additionally, \(E_J\) and \(C_J\) denote the Josephson energy and the capacitance of the SQUID, respectively. As we anticipated, the description of the cavity involves a field \(\Phi(x,t)\) for \(0<x<L_0\) and the additional degree of freedom \(\Phi(L_0,t)\). The dynamic equations are
\begin{equation}
\partial_t^2 \Phi - v^2 \partial_x^2 \Phi = 0
\,,
\end{equation}
(from now on we will take \(v=1\)), and~\cite{lomb.numerico1, lomb.numerico2}
\begin{equation}
\label{lomb.chap:repaso-eq:phid}
\frac{\hbar^2}{E_C} \partial_t^2 \Phi(L_0,t) + 2 E_J \cos{f(t)} \Phi(L_0,t) + E_{l cav} L_0 \partial_x \Phi(L_0,t)  = 0
\end{equation}
where \(E_C = (2e)^2/(2 C_J)\) and \(E_{l cav} = (\hbar/2e)^2 (1/l L_0)\). The above equation arises from the variation of the action with respect to \(\Phi(L_0,t)\) and can be considered as a generalized boundary condition for the field. We could also ask for generalized boundary conditions at \(x=0\), but for simplicity, we shall assume that \(\Phi(0,t)=0\) (which physically corresponds to the situation where the cavity is decoupled). Under certain choices of the cavity and SQUID parameters, and by adjusting the magnetic field through the SQUID, the second time derivative of the field becomes negligible, and Eq.~\eqref{lomb.chap:repaso-eq:phid} can be written as

\begin{equation}
    0 = \Phi(L_0,t) + \frac{E_{l cav} L_0}{2 E_J \cos{f(t)}} \partial_x \Phi(L_0,t)
    \approx \Phi\left(L_0+\frac{E_{l cav} L_0}{2 E_J \cos{f(t)}},t\right)
\end{equation}
which implies that the superconducting cavity behaves like a perfect cavity with a moving mirror at the end, meaning that \(L_0\) effectively becomes a function of time denoted as \(L_0 \equiv L(t)=L_0+\frac{E_{l cav} L_0}{2 E_J \cos{f(t)}}\). In this scenario, the boundary conditions can be written as \(\Phi(0,t)=\Phi(L(t),t)=0\)~\cite{lomb.PRE-paula,lomb.PRA-paula-2espejos,lomb.louko-super}.

In fact, the effects of non-stationary boundary conditions have been observed in a superconducting circuit~\cite{lomb.zeilinger,lomb.dce_observacion}. In particular, in the work~\cite{lomb.cruz_2022}, we have studied how SQUIDs located in the middle and at the ends of a superconducting waveguide allow the simulation of an optomechanical system formed by two perfectly reflecting walls and a dielectric membrane between them with arbitrary permittivity separating both halves. Thus, these more general boundary conditions allow us to adjust both the spectrum of the cavity and the shape of its electromagnetic modes.

In~\cite{lomb.numerico1} we have presented a detailed numerical analysis of the particle creation for a quantum field in the presence of boundary conditions that involve a time-dependent linear combination of the field and its spatial and time derivatives. We have evaluated numerically the Bogoliubov transformation between in and out-states and found that the rate of particle production strongly depends on whether the spectrum of the unperturbed cavity is equidistant or not, and also on the amplitude of the temporal oscillations of the boundary conditions. We have provided some analytical justifications, based on MSA, for
the different regimes found numerically and emphasized the dependence of the results on the main characteristics of the spectrum.

Furthermore, in~\cite{lomb.numerico2} we have presented an analytical and numerical analysis of the particle creation in a tunable cavity ended with two SQUIDs, both subjected to external time-dependent magnetic fields. We considered a situation in which the boundary conditions at both ends are periodic functions of time.
The boundary conditions for this case are:
\begin{equation}\label{lomb.eqphi0}
\frac{\hbar^2}{E_C} \ddot \phi_0 + 2 E^L_J \cos{f^L(t)} \phi_0 + E_{\rm l,cav }d \phi'_0  = 0
\,,
\end{equation}
at $x = 0$ and
\begin{equation}\label{lomb.eqphid}
\frac{\hbar^2}{E_C} \ddot \phi_d + 2 E^R_J \cos{f^R(t)} \phi_d + E_{\rm l,cav} d \phi'_d  = 0
\,,
\end{equation}
at $x = d$. The spectrum is determined by the solution of that system of equations, that can be rewritten in terms of the new parameters of the cavity $\chi_0$ and $b_{0L,R}$ as
\begin{eqnarray}
\label{lomb.spectrum2}
(k_n d)\tan{(k_n d + \varphi_n)}  + \chi_0 (k_n d)^2 &=&  b_{0R}  \nonumber \\
-(k_n d)\tan{\varphi_n} + \chi_0  (k_n d)^2  &=&  b_{0L} ,
\end{eqnarray}
where we have set $b_{0L,R} = V_0^{L,R} \cos f^{L,R}_0$ with $V_0^{L,R}= 2 E_J^{L,R}/E_{\rm l,cav}$, and $\chi_0 = 2 C_J/(C_0 d)$.
The three free parameters that determine the solutions of Eq.~(\ref{lomb.spectrum2}) are $\chi_0$, $b_{0R,L}$.

We have shown that there is parametric resonance when the external frequencies are of the form \(\Omega_{L,R} = 2k_n\) and/or \(\Omega_{L,R} = k_n \pm k_m\) where \(k_n\) and \(k_m\) are eigenfrequencies of the static cavity. Under parametric resonance, the number of created particles grows exponentially, with a rate that depends not only on the amplitudes and frequencies of the external modulations but also on the parameters of the static cavity. Moreover, the relative phase of the external modulation introduces interference effects in the rate of growth, in the sense that the number of created photons when two SQUIDs are externally pumped is not the sum of the created particles by each individually pumped SQUID.

We have further investigated regimes that are not reachable with the lowest order MSA. For equal driving frequencies ($\Omega_R = \Omega_L$) and large values of the parameter $b_0$, we showed that the particle creation rate grows quadratically with the final time for breathing modes and that particle creation is suppressed in the translational modes. On the other hand, by setting the parameters of the static cavity so that the spectrum becomes non-equidistant, we found exponential rates for particle creation. In this case, we also found interference effects, describing situations in which the destructive interference is total (no exponential growth in the translational modes) and cases where it is partial (exponential growth with different rates in both breathing and translational modes). The amount of interference can be tuned by adjusting the static magnetic fluxes on the SQUIDs. We obtained similar results when the external frequencies are different and found exponential growth of the number of created particles not only for the usual case where the frequencies are twice an eigenfrequency of the static cavity but also when given by the sum of two modes $k_m + k_n$.

Therefore, in conclusion, in Refs.~\cite{lomb.numerico1,lomb.numerico2,lomb.PRE-paula,lomb.PRA-paula-2espejos,lomb.cruz_2022}, we have performed a comprehensive analysis of spectra, resonances, and photon generation across various geometries and electromagnetic boundary conditions.


\section{Quantum thermodynamics and DCE in superconducting cavities}
\label{lomb.chap:otto}

Following~\cite{lomb.Otto_nos}, in this section we will introduce the fundamental concepts of quantum thermodynamics and apply them to study a heat engine consisting of a quantum field in a cavity undergoing a quantum Otto cycle. The goal is to understand the impact of the DCE on the efficiency and power output of a quantum heat engine.

In general terms, the quantum Otto cycle involves a system, or working medium, governed by a Hamiltonian \( H_0 \) to which four basic operations are applied. First, the system is put in contact with a cold bath at inverse temperature \( \beta_A \); this leaves the system in a thermal state with internal energy \( E_A \). Second, the system is isolated from the bath and subjected to a time-dependent Hamiltonian, reaching a state with internal energy \( E_B \). Third, the system is put in contact with a hot bath at inverse temperature \( \beta_C \), reaching another thermal state, but this time with internal energy \( E_C \). Finally, the system is once again isolated from the bath and subjected to a time-dependent Hamiltonian that restores the original Hamiltonian \( H_0 \), leaving the system in a state with internal energy \( E_D \).

As a result, in the third branch, the machine receives heat \( Q = E_C - E_B \) and work is extracted between operations A and B, given by \( W_{AB} = E_A - E_B \), and between C and D, \( W_{CD} = E_C - E_D \). This sums to a net extracted work of \( W = W_{AB} + W_{CD} \). If this work is positive, \( W > 0 \), we say the machine acts as a heat engine, meaning it converts heat from thermal baths into useful work with an efficiency given by
$\eta = \frac{W}{Q}$.

On the other hand, if the work and heat are negative, \( W, Q < 0 \), the machine is said to act as a refrigerator, meaning it consumes work and uses it to heat a hot reservoir and cool the cold one.

Specifically, we will study the following implementation of an Otto cycle for a quantum scalar field in an optomechanical cavity with a moving wall (although it can also be implemented in cQED architectures). If the cavity wall describes a trajectory between \( t = 0 \) and \( t = \tau \) given by
\begin{equation}
\label{lomb.chap:maquina-eq:L}
    L(t) = L_0 + \delta L(t) = L_0[1 - \epsilon \delta(t)]
\end{equation}
with \( 0 = \delta(0) < \delta(t) < \delta(\tau) = 1 \) and \( 0 < \epsilon \ll 1 \), then the quantum field Hamiltonian can be written as
\begin{equation}
\label{lomb.chap:maquina-eq:H}
    H(t) = H_0 + H_1(t)
\end{equation}
where the Hamiltonian \( H_0 \) can be decomposed as
\begin{equation}
    H_0 = H_{\text{free}} + E_{\rm SC}(L),
\end{equation}
where \( H_{\text{free}} = \sum_{k=1}^\infty \hbar \omega_{k} N_k \) is the free evolution term of the modes with photon number operator \( N_k = a_k^\dagger a_k \) in mode frequency \( \omega_{k} = \omega_k(L_0) \), while \( E_{\rm SC}(L_0) = -\pi \hbar / (24 L_0) \) is the energy associated with the static Casimir effect. On the other hand, the dynamic part of the wall's evolution is given by Eq.~(\ref{lomb.chap:repaso-eq})
\begin{align}
\label{lomb.chap:maquina-eq:H1}
    H_1(t) &= \hbar \sum_{k=1}^{\infty} \left[ \omega_{k}^{\prime} \delta L(t) a_{k}^{\dagger} a_{k} + \frac{\omega_{k}^{\prime}}{2} \delta L(t) \left( a_{k}^{\dagger 2} + a_{k}^{2} \right) \right] \nonumber\\
    &\quad + \frac{\hbar}{2i} \sum_{k,j=1}^{\infty} \frac{\dot{\delta L}(t)}{L_0} g_{kj} \sqrt{\frac{\omega_{k}}{\omega_{j}}} \left( a_{k} a_{j} - a_{k}^{\dagger} a_{j} + a_{k} a_{j}^{\dagger} - a_{k}^{\dagger} a_{j}^{\dagger} \right),
\end{align}
an expression valid to first order in \( \epsilon \) where the coupling constants \( g_{kj} \) are given by
\begin{equation}
    g_{kj} = -g_{jk} = L \int_{0}^{L} dx \left( \partial_{L} \psi_{k} \right) \psi_{j},
\end{equation}
essentially the same constants \( M_{kj} \) from Sec.~\ref{lomb.chap:DCEcQED} made dimensionless by length.

\begin{figure}[!htb]
\begin{center}
\includegraphics[width=0.60\textwidth]{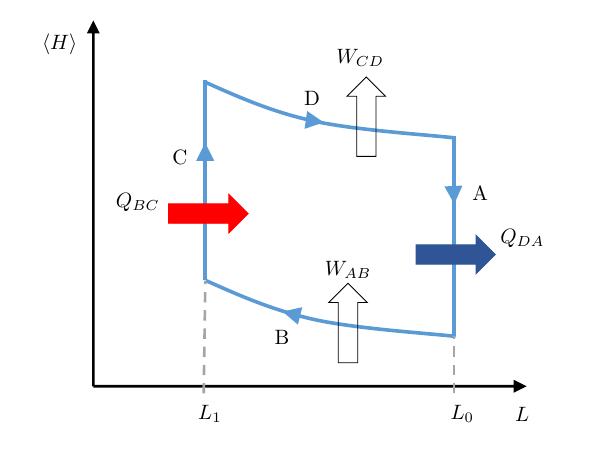} {}
\end{center}
\caption{Otto cycle in terms of cavity length and the average value of energy in the cavity.}
\label{lomb.ciclo}
\end{figure}

Thus, the cycle is represented in Fig.~\ref{lomb.ciclo} and can be described as: first, the system is put in contact with the cold bath at inverse temperature \( \beta_A \). We assume perfect thermalization, leaving the system in a thermal state
    \begin{equation}
        \rho^{\beta_A} = \frac{\exp(-\beta_A H_{\text{free}})}{Z} = \prod_{k=1}^\infty \frac{e^{-\beta_A \hbar \omega_k N_k}}{Z_k},
    \end{equation}
    with \( Z_k = \text{Tr}(e^{-\beta_A \hbar \omega_k N_k}) = \frac{1}{1 - e^{-\beta_A \hbar \omega_k}} \) and internal energy:

    \begin{eqnarray}
        E_A &=& \text{Tr}(\rho^{\beta_A} \frac{H_{\text{free}}}{Z}) + E_{\rm SC}(L_0)
        = \sum_{k=1}^\infty \frac{\hbar \omega_k}{e^{\beta_A \hbar \omega_k} - 1} + E_{\rm SC}(L_0) \nonumber\\
        &\equiv& \sum_{k=1}^\infty \hbar \omega_k \bar{N}_k^{\beta_A} + E_{\rm SC}(L_0),
    \end{eqnarray}
    where \( \bar{N}_k^{\beta_A} \) is the thermal occupation number.

 Second, the wall is displaced from \( L_0 \) to \( L_1 = L(\tau) = L_0(1 - \epsilon) \), following trajectory Eq. (\ref{lomb.chap:maquina-eq:L}) and compressing the cavity. The system evolves under Hamiltonian Eq. (\ref{lomb.chap:maquina-eq:H1}), resulting in internal energy
    \begin{equation}
        E_B = \sum_{k=1}^{\infty} \hbar \omega_k(L_1) \text{Tr}(\rho N_k) + E_{\rm SC}(L_1).
    \end{equation}

    Third, the system is put in contact with a hot bath at inverse temperature \( \beta_C \), leaving it in a thermal state with internal energy
    \begin{align}
        E_C = \sum_{k=1}^\infty \hbar \omega_k(L_1) \bar{N}_k^{\beta_C} + E_{\rm SC}(L_1),
    \end{align}
    where \( \bar{N}_k^{\beta_C} \) is evaluated at \( \omega_k(L_1) \).

    Fourth, the wall is moved back from \( L_1 \) to \( L_0 \), following the time-reversed trajectory \( \tilde{L}(t) = L(\tau - t) \). Thus, the cavity expands to its original size, ending the process with internal energy
    \begin{equation}
        E_D = \sum_{k=1}^{\infty} \hbar \omega_k \text{Tr}(\tilde{\rho} N_k) + E_{\rm SC}(L_0),
    \end{equation}
where \( \tilde{\rho} \) is the density matrix associated with the reverse path.

\subsection{Adiabatic evolution}

If we now assume the wall moves slowly enough to satisfy the quantum adiabatic theorem conditions, i.e., if \( \tau \gg 1 / \omega_1 \), there will be no change in the energy level populations, giving
\begin{eqnarray}
	\text{Tr}({\rho N_k}) &=& \text{Tr}({\rho^{\beta_A} N_k}) \\
	\text{Tr}({\tilde{\rho} N_k}) &=& \text{Tr}({\rho^{\beta_C} N_k}).
\end{eqnarray}
Then, the heat supplied to the system would simply be the difference
\begin{align}
	Q^{\rm Otto} = E_C - E_B = \sum_{k=1}^\infty \hbar \omega_k(L_1)(\bar{N}_k^{\beta_C} - \bar{N}_k^{\beta_A}),
\end{align}
and the work is given by
\begin{equation}
W^{\rm Otto} = (E_A - E_B) + (E_C - E_D)
= \sum_{k=1}^\infty \hbar (\omega_k(L_1) - \omega_k)(\bar{N}_k^{\beta_C} - \bar{N}_k^{\beta_A}).
\end{equation}
As we can see, the static Casimir energy \( E_{\rm SC} \) does not affect the heat or work provided by the heat engine. Additionally, if the condition
\[
    \omega_k \beta_A \leq \omega_k(L_1) \beta_C
\]
is satisfied for all \( k \), then \( Q^{\rm Otto} > 0 \). We also see that since \( \omega_k(L_1) > \omega_k \) for \( L_1 < L_0 \), the work ends up being positive \( W > 0 \).

In light of these results, the efficiency can be written as
\begin{align}
\eta^{\rm Otto} &= \frac{W^{\rm Otto}}{Q^{\rm Otto}}
= \frac{\sum_{k=1}^\infty \hbar (\omega_k(L_1) - \omega_k)(\bar{N}_k^{\beta_C} - \bar{N}_k^{\beta_A})}{\sum_{k=1}^\infty \hbar \omega_k(L_1)(\bar{N}_k^{\beta_C} - \bar{N}_k^{\beta_A})} \nonumber \\
&= \frac{\sum_{k=1}^\infty \hbar \omega_k(L_1)(1 - \omega_k / \omega_k(L_1))(\bar{N}_k^{\beta_C} - \bar{N}_k^{\beta_A})}{\sum_{k=1}^\infty \hbar \omega_k(L_1)(\bar{N}_k^{\beta_C} - \bar{N}_k^{\beta_A})}.
\end{align}
And if the spectrum is given by \( \omega_k(L) = k \pi / L \), we have
\begin{equation}
	\frac{\omega_k}{\omega_k(L_1)} = \frac{k \pi / L_0}{k \pi / L_1} = \frac{L_1}{L_0} = 1 - \epsilon,
\end{equation}
from which we derive the following simple result for efficiency $\eta^{\rm Otto} = \epsilon$.

This expression implies that the efficiency of the Otto cycle for this system, in the adiabatic limit, depends only on the compression ratio. Furthermore, we should highlight that if we want to achieve the maximum possible efficiency (which is the Carnot cycle efficiency \( \eta^{\rm Carnot} \)), then the thermal baths and cavity must satisfy
\begin{equation}
	\epsilon = \eta^{\rm Otto} = \eta^{\rm Carnot} = 1 - \frac{\beta_C}{\beta_A}.
\end{equation}

\subsection{Non-adiabatic evolution}

In this section, we will go beyond the adiabatic approximation and show that the DCE induces a type of quantum friction that reduces the efficiency of the cycle. To do this, we need to calculate the energy after a non-adiabatic movement of the wall
\begin{align}
E=\text{Tr}({\rho H})=\text{Tr}(U{\rho^{\beta_A}U^\dagger H}),
\end{align}
where \( U \) is the unitary operation that induces the wall’s movement on the field state.

To calculate this magnitude, we will proceed using perturbation theory to the lowest order in non-adiabatic epsilon.
Our Hamiltonian, given by Eq.~(\ref{lomb.chap:maquina-eq:H}), is suitable for this calculation in the interaction representation. It is important to remember that, in this representation, the states evolve according to
\begin{equation}
	|\psi_I(t)\rangle=U_1|\psi_S\rangle
\end{equation}
where \( U_1 \) is given by
\[
U_1 = \mathcal{T} \exp\left(-\frac{i}{\hbar} \int_0^t H_1(t) \, dt \right),
\]
where \( \mathcal{T} \) indicates time-ordering. In this representation, operators vary with time as
\[
A_I(t) = U_0 A_S(t) U_0^\dagger,
\]
where \( U_0 = \exp\left(-\frac{i}{\hbar} H_{\text{free}} t \right) \). Here, the subscripts \( I \) and \( S \) refer to the interaction and Schrödinger representations, respectively.

In our case, the energy of the quantum field at time \( t = \tau \) is given by
\begin{align}
\label{lomb.chap:maquina-eq:energia1}
E(\tau) = \text{Tr}({\rho^{\beta_A} U_1^\dagger H_{I}(\tau) U_1}).
\end{align}

We will calculate this energy perturbatively up to second order in \( \epsilon \). We can do this by first approximating \( U_1 \) to second order and substituting that result into Eq.~(\ref{lomb.chap:maquina-eq:energia1}) to obtain
\begin{eqnarray}
\label{lomb.chap:maquina-eq:Napprox}
E(\tau) = &&\text{Tr}(\rho^{\beta_A} H_{I}(\tau)) - \frac{1}{\hbar} i \int_{0}^{\tau} dt_{1} [\text{Tr}(\rho^{\beta_A} H_I H_{1,I}(t_{1})) - \text{Tr}(\rho^{\beta_A} H_{1,I}(t_{1}) H_I)] \nonumber \\
&& + \frac{1}{\hbar^{2}} (-i)^{2} \int_{0}^{\tau} dt_{1} \int_{0}^{\tau} dt_{2} \text{Tr}(\rho^{\beta_A} H_{1,I}(t_{1}) H_I H_{1,I}(t_{2})) \nonumber \\
&& + \frac{1}{\hbar^{2}} (-i)^{2} \int_{0}^{\tau} dt_{1} \int_{0}^{t_{1}} dt_{2} \text{Tr}(\rho^{\beta_A} H_{1,I}(t_{2}) H_{1,I}(t_{1}) H_I) \nonumber \\
&& + \frac{1}{\hbar^{2}} (i)^{2} \int_{0}^{\tau} dt_{1} \int_{0}^{t_{1}} dt_{2} \text{Tr}(\rho^{\beta_A} H_I H_{1,I}(t_{1}) H_{1,I}(t_{2})).
\end{eqnarray}

As the Hamiltonian was expressed to first order in \( \epsilon \) and we seek the energy to second order, we need to expand the bosonic operators according to Eq.~(\ref{lomb.chap:primerorden}) to obtain the following expansion of the terms of \( H_0 \) in the interaction representation:
\begin{equation}
\omega_k(L) N_k(\omega(L)) =  \omega_k(1 + \delta L(t)) [N_k + \delta L \frac{\omega_{k}^{\prime}}{2\omega_{k}} (e^{i2\omega_k t} a_k^{\dagger 2} + e^{-i2\omega_k t} a_k^{2})
 + \left(\delta L \frac{\omega_{k}^{\prime}}{2\omega_{k}}\right)^{2} (2N_k + 1)].
\label{lomb.eq:24}
\end{equation}

Replacing this result in Eq.~(\ref{lomb.chap:maquina-eq:Napprox}) and simplifying the expression, we find that the internal energy is given by
\begin{eqnarray}
	E(\tau) &=& \sum_{k} \bigg[ \hbar \omega_{k}(\tau) \bar{N}_{k}^{\beta_A} + \frac{\epsilon^2}{4} \hbar \omega_{k} \int_{0}^{\tau} dt_{1} \int_{0}^{\tau} dt_{2} F^{\beta_A}(t_{1}, t_{2}) \bigg].
\end{eqnarray}
where
\begin{eqnarray}
\label{lomb.chap:maquina-eq:F}
	&&F^{\beta_A}(t_{1}, t_{2}) = \frac{\omega_{k}^{\prime 2} L_0^2}{\omega_{k}^{2}} \dot{\delta}(t_1) \dot{\delta}(t_2) \cos\left[ 2\omega_{k}(t_{1} - t_{2}) \right] \left\{ 2\bar{N}_{k}^{\beta_A} + 1 \right\} \nonumber \\
	&& + \sum_{j=0}^{\infty} \dot{\delta}(t_1) \dot{\delta}(t_2) \frac{g_{jk}^{2}}{\omega_{j}\omega_{k}} \bigg[ (\omega_{k} - \omega_{j})^{2} \cos\left[ (\omega_{j} + \omega_{k})(t_{1} - t_{2}) \right] \left\{ \bar{N}_{k}^{\beta_A} + \bar{N}_{j}^{\beta_A} + 1 \right\} \nonumber \\
	&& + \left( \omega_{j} + \omega_{k} \right)^{2} \cos\left[ (\omega_{j} - \omega_{k})(t_{1} - t_{2}) \right] \left\{ \bar{N}_{j}^{\beta_A} - \bar{N}_{k}^{\beta_A} \right\} \bigg].
\end{eqnarray}

We can see that the non-adiabatic contribution to the energy is
\begin{align}
\label{lomb.chap:maquina-eq:frictionEnergy}
E_F^{\beta_A}(\tau) = \frac{\epsilon^2}{4} \sum_{k} \hbar \omega_{k} \int_{0}^{\tau} dt_{1} \int_{0}^{\tau} dt_{2} F^{\beta_A}(t_{1}, t_{2})\, ,
\end{align}
which is a form of quantum friction~\cite{lomb.salamon, lomb.ottoion, lomb.stefanatos, lomb.stefanatos2, lomb.otto_kosloff}. This is because it is not conservative, as it is not a function of the state of the cavity, but fundamentally depends on the trajectory \( \delta(t) \) used to reach the final state. Furthermore, since \( E_F \) depends quadratically on \( \delta(t) \) and \( \dot{\delta}(t) \), this energy contribution will be the same in magnitude and sign, regardless of whether the wall moves forward or backward. This is a clear contrast to conservative forces and more similar to the energy dissipated to a viscous medium, which depends on the trajectory (and even the speed \( \dot{\delta} \)) but not on the direction.

Moreover, we can show that this energy is always positively defined as the third term of the same equation can be written as
\begin{align}
\sum_{k,j=1}^{\infty} \frac{h_{jk}}{\omega_{j}} \left[ \bar{N}_{j}^{\beta_A} - \bar{N}_{k}^{\beta_A} \right] = \sum_{k>j=1}^{\infty} \frac{h_{jk}}{\omega_{j} \omega_{k}} (\omega_{k} - \omega_{j}) \left[ \bar{N}_{j}^{\beta_A} - \bar{N}_{k}^{\beta_A} \right] \nonumber,
\end{align}
which is positive because \( \omega_{k} > \omega_{j} \) for \( k > j \), and hence \( \bar{N}_{j}^{\beta_A} \geq \bar{N}_{k}^{\beta_A} \). This proves that
\begin{align}
E_F^{\beta_A}(\tau) \geq 0,
\end{align}
as we wanted to show.

We have also established an upper bound for the energy lost due to quantum friction by assuming that the wall has zero acceleration at the beginning and at the end of the motion, $\ddot{\delta}(0)=\ddot{\delta}(\tau)=0$. In this way, by making a few reasonable assumptions, we can bound the frictional energy due to the DCE over a wide range of trajectories~\cite{lomb.Otto_nos}.

\subsection{Quantum thermal engine}

After defining the system, the cycle to be performed, and calculating the energy at the end of each branch in the non-adiabatic limit, we are in a position to analyze the efficiency of this machine used as a thermal engine. To do this, by taking the energy generated by the moving wall, it is straightforward to calculate the dissipated heat.
\begin{align}
Q&= E_{C}-E_{B}\nonumber\\
&=\sum_{k=1}^{\infty}\hbar\omega_{k}(L_{1})\bar{N}_{k}^{\beta_{C}}-\sum_{k=1}^\infty\bigg[\hbar\omega_{k}(L_{1})\bar{N}_{k}^{\beta_{A}}+\epsilon^2\frac{\hbar\omega_{k}(L_1)}{4}\int_{0}^{\tau}dt_{1}\int_{0}^{\tau}dt_{2}F^{\beta_{A}}(t_{1},t_{2})\bigg].
\end{align}
Given that the frictional energy is already quadratic in $\epsilon$, we can approximate $\omega_{k}(L_{1})\approx\omega_{k}$ in that term, and thus the heat is given by
\begin{align}
Q&=Q^{\rm Otto}-E_{F}^{\beta_{C}}(\tau),
\end{align}
up to second order in $\epsilon$.

The work is also easy to calculate
\begin{equation}
	W =W_{AB}+W_{CD}
	=W^{\rm Otto}-\left[E_{F}^{\beta_{A}}+E_{F}^{\beta_{C}}\right].
\end{equation}
Finally, the efficiency of the thermal engine for the non-adiabatic cycle is given by
\begin{align}
	\eta &= \frac{W^{\rm Otto}-\left[E_{F}^{\beta_{A}}+E_{F}^{\beta_{C}}\right]}{Q^{\rm Otto}-E_{F}^{\beta_{C}}}
	\approx\frac{W^{\rm Otto}}{Q^{\rm Otto}}+\frac{W^{\rm Otto}E_{F}^{\beta_{C}}-Q^{\rm Otto}\left[E_{F}^{\beta_{A}}+E_{F}^{\beta_{C}}\right]}{(Q^{\rm Otto})^{2}}+\mathcal{O}(\epsilon^{3})\nonumber\\
	&\approx\eta^{\rm Otto}-\frac{E_{F}^{\beta_{A}}+E_{F}^{\beta_{C}}}{Q^{\rm Otto}}+\mathcal{O}(\epsilon^{3}),
\end{align}
which is always less than the adiabatic efficiency $\eta^{\rm Otto}$.

\subsection{DCE as quantum friction energy
}\label{lomb.chap:maquina-subsec:Efric}

\begin{figure}[!htb]
\begin{center}
\includegraphics[width=0.57\textwidth]{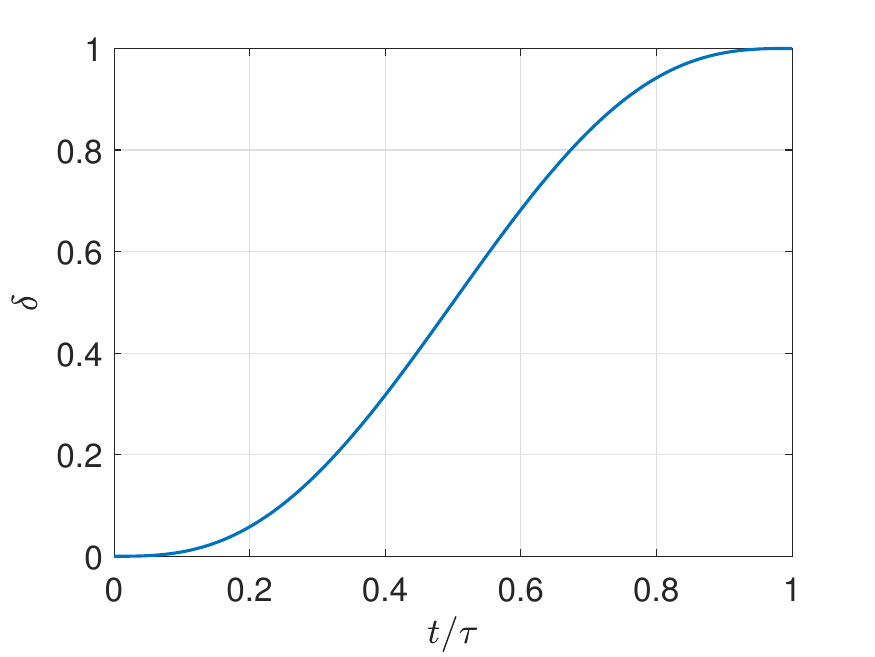} {}
\end{center}
\caption{Trajectory to exemplify the results, which satisfies all the constraints imposed.}
\label{lomb.chap:maquina-fig:tray}
\end{figure}

\begin{figure}[!htb]
\begin{center}
\includegraphics[width=0.60\textwidth]{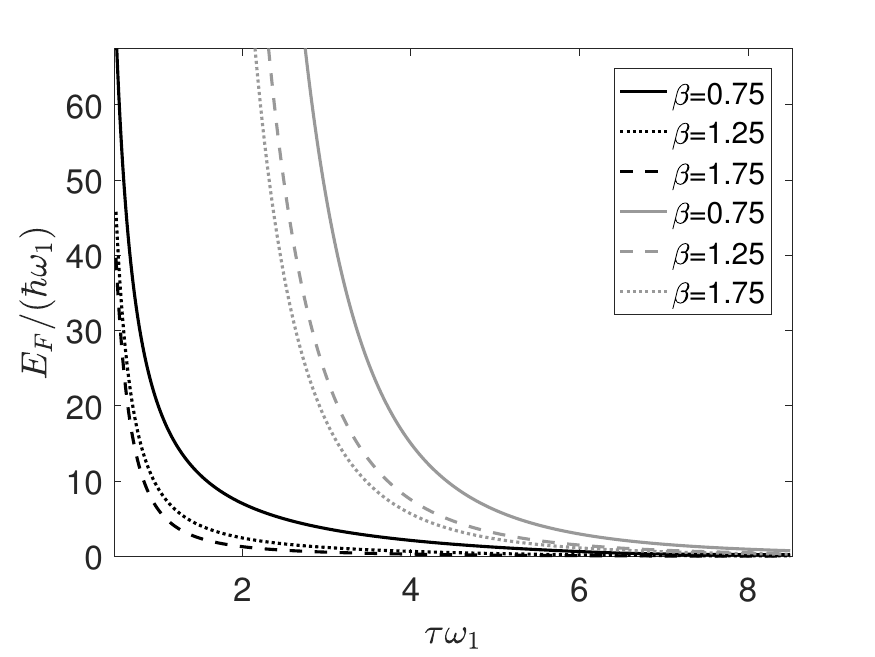} {}
\end{center}
\caption{Friction energy as a function of the dimensionless time for different values of $\beta$. The computed friction energy for the example trajectory is in black lines. We set $\epsilon=0.01$.
}
\label{lomb.chap:maquina-fig:E_t}
\end{figure}

To illustrate our results, we consider the trajectory given by the lowest-order polynomial $\delta(t)=10(t/\tau)^3-15(t/\tau)^4+6(t/\tau)^5$,
that satisfies all the following conditions $\delta(0)=\dot{\delta}(0)=\ddot{\delta}(0)=0$, $\dot{\delta}(\tau)=\ddot{\delta}(\tau)=0$, and $\delta(\tau)=1$.
This trajectory is shown in Fig.~(\ref{lomb.chap:maquina-fig:tray}). Using this trajectory and the previous results, we calculate the quantum friction energy produced by the DCE, as shown in Fig.~(\ref{lomb.chap:maquina-fig:E_t}). It is important to note that this energy becomes zero for very slow motions ($\tau\to\infty$) and becomes arbitrarily large for increasingly rapid movements of the wall. On the other hand, for a fixed $\tau$, the friction energy converges to a finite value as the inverse temperature of the initial state $\beta$ is increased. This is clearly associated with the photon production given by the DCE, which persists even when starting from a vacuum state.

\begin{figure}[!htb]
\begin{center}
\includegraphics[width=0.60\textwidth]{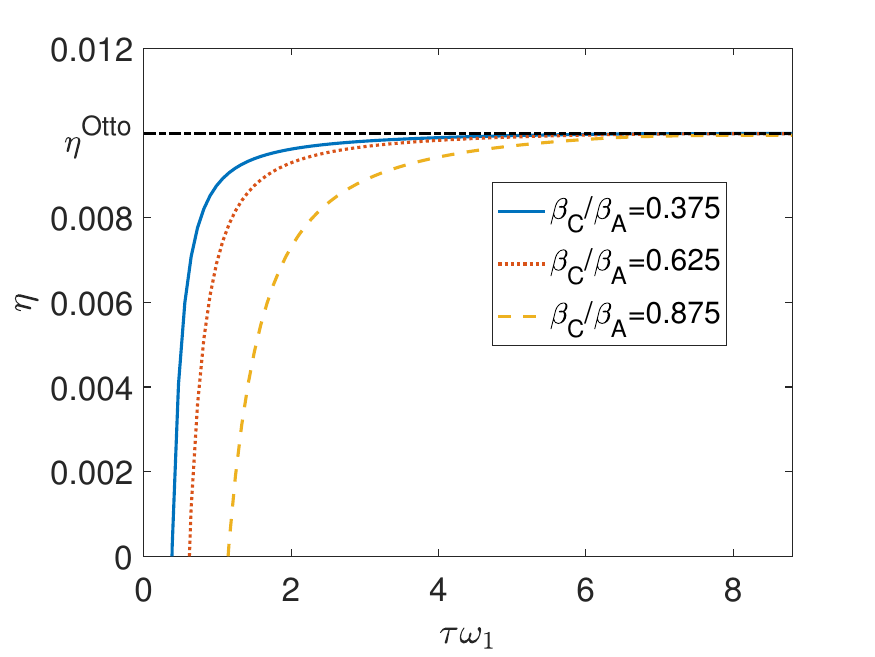} {}
\end{center}
\caption{Efficiency of an engine for different ratios between the temperatures of the thermal baths ($\beta_C/\beta_A$) as a function of the dimensionless timescale $\tau\omega_1$. We set  $\epsilon=0.01$.}
\label{lomb.chap:maquina-fig:eta}
\end{figure}

As we mentioned earlier, the friction energy from the DCE is always positive and, similar to friction in a classical piston, decreases the efficiency of the engine compared to the efficiency of the non-adiabatic Otto cycle $\eta^{\rm Otto}$. In general, quantum friction increases as the movement of the wall becomes faster ($\tau\to0$). In fact, due to this, a minimum timescale arises below which the engine can no longer function as such, as the work delivered becomes negative (see Fig.~\ref{lomb.chap:maquina-fig:eta}).

\begin{figure}[!htb]
\begin{center}
\includegraphics[width=0.60\textwidth]{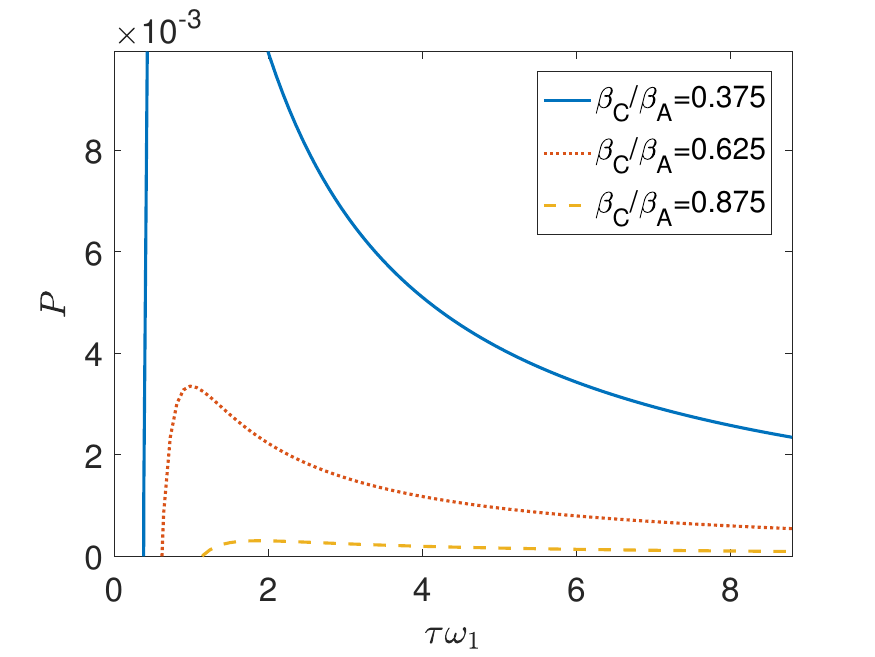} {}
\end{center}
\caption{Power produced  for different temperature ratios as a function of time. There is a peak at approximately $\tau\omega_1\sim1$, indicating an optimal timescale for operation. We set $\epsilon=0.01$.}
\label{lomb.chap:maquina-fig:chap5-P}
\end{figure}

We continue the study of the power $P$ produced by this thermal engine. In the adiabatic case, when $\tau\omega_1\gg 1$, the work is independent of the timescale $\tau$, and thus the power increases as time decreases, given that $P\sim 1/\tau$. In the opposite limit $\tau\omega_1\ll 1$ (non-adiabatic), the power turns out to be proportional to $1/\tau^4$. Therefore, the power will have two contributions, with different signs, that scale with different powers of $\tau$. Consequently, we expect a peak around the time when the friction energy begins to be comparable to that obtained in the adiabatic limit, $\tau\omega_1\sim1$. This is illustrated in Fig.~\ref{lomb.chap:maquina-fig:chap5-P}.

Finally, in~\cite{lomb.Otto_nos} it has been shown that the work extracted is positive for large temperature ratios between the baths ($\beta_C/\beta_A$) and long timescales. Conversely, if $\tau\rightarrow 0$ and $\beta_C/\beta_A\rightarrow 1$, it is observed that the work becomes zero.

\section{Controlled-Squeeze gate}
\label{lomb.chap:ctlsqz}

As we mentioned above, circuit quantum electrodynamics has become the leading architecture for quantum computation.
Circuit QED has already been used to manipulate tens of qubits for quantum simulation~\cite{lomb.houck2012chip,lomb.andersen2024thermalization} and quantum error correction~\cite{lomb.reed2012realization}.
Even at the small scale of a few resonators, cQED provides an alternative to Cavity QED~\cite{lomb.haroche2020cavity} with practical advantages. It has been used to study the dynamical Casimir effect (DCE)~\cite{lomb.dce_observacion,lomb.dyncasimir2}, foundational aspects of quantum mechanics~\cite{lomb.Devoret1985,lomb.minev2019},  and also enabled practical applications like error correction of non-classical states in resonators~\cite{lomb.ofek2016,lomb.sivak2022} and quantum communication~\cite{lomb.storz2023loophole}.

The preparation and control of non-classical quantum states of the electromagnetic field in the resonator is crucial to perform universal simulations which, in principle, involve the manipulation of arbitrary quantum states.

To generate arbitrary states of the cavity field with limited resources, it is crucial to identify a universal set of gates~\cite{lomb.khaneja2005optimal,lomb.leghtas2013deterministic,lomb.heeres2017implementing,lomb.Krastanov2015, lomb.Kundra2022}. In this context, it has been shown that arbitrary unitary operators acting on the resonator field states can be generated using Gaussian operations together with controlled displacement operators~\cite{lomb.eickbusch2022} (which are denoted as $\textbf{C-Dsp}(\gamma)$ and apply a displacement $\hat{D}(\gamma)$ depending on the state of a control qubit). Complemented with Gaussian operations on the field, this controlled gate provides a universal resource to create arbitrary states.

In this section, we present another gate, based on controlled squeezing, that provides a universal resource, and show that it demands reasonable experimental resources. We will denote it as $\textbf{C-Sqz}(r,\theta)$ as it applies the squeezing operation $\hat{S}(r,\theta)$, conditioned on the state of a control qubit. We also present an example protocol to make use of this gate in a cQED setup to encode quantum states in the resonator using an encoding that makes the errors induced by the loss of a photon detectable through parity measurement. But before going into this, it is important to note that the proposed gate is universal, as it was shown in Ref.~\cite{lomb.ctrlsqz}.
We proved the universality of the controlled-squeeze gate using the universality of the controlled-displacement $\textbf{C-Dsp}(\gamma)$.
Our proposal enables the implementation of controlled displacements and controlled squeezing gates in the same setup, and therefore adds flexibility to the cQED architecture, simplifying certain tasks.

\begin{figure}
	\centering	\includegraphics[width=0.7\textwidth]{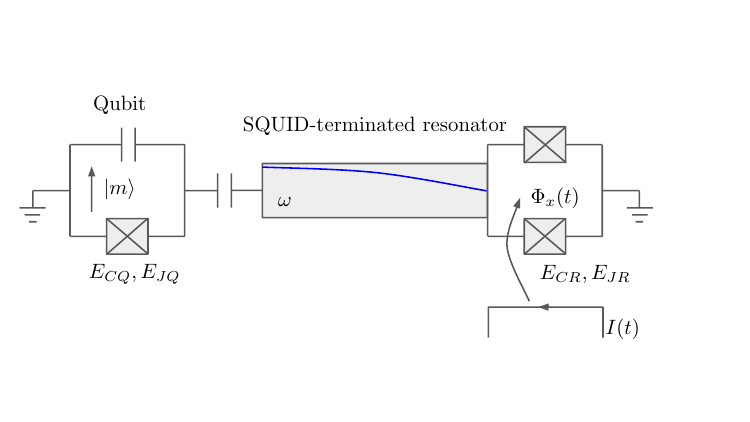}
		\caption{Setup: coplanar wave guide terminated in a SQUID at the right and capacitively coupled to a transmon at the left.  The transmon is initially prepared in a state $|m\rangle$ (where $m=0,1$) and acts as a switch that either tunes or detunes the resonator to the driving frequency.}
		\label{lomb.chap:setup}
\end{figure}

To implement the proposed controlled-squeeze gate we use two basic elements: one frequency-tunable resonator and a qubit~\cite{lomb.Paik2011,lomb.wang2022towards} with quantum states $\vert 0\rangle$ and $\vert 1\rangle$. This qubit is dispersively coupled to the resonator in the number splitting regime, so that its frequency depends on the state of the qubit (i.e. it is $\bar\omega_0$ if the state is $\vert 0\rangle$ and $\bar\omega_1$ if the state is $\vert 1\rangle$).
The resonator is terminated by a SQUID, where we apply a time-dependent flux $\Phi(t) = \epsilon\sin(\omega_d t)$. In this way the resonator's natural frequency becomes a time-dependent parameter $ \omega_{0,1}(t)=\bar\omega_{0,1}+ g_d\epsilon  \sin( \omega_d t-\theta)$,  where $g_d$ is the driving coupling constant between the SQUID and the resonator modes (see below) and  $\epsilon$ is the amplitude of the flux drive. When the driving frequency is $\omega_d = 2 \bar\omega_1$, then parametric resonance takes place and the cavity's field is squeezed as a result~\cite{lomb.squeeze_states_param}. If the detuning $\Delta = \bar\omega_1- \bar\omega_0$ is $\Delta\gg g_d \epsilon$ then the state of the cavity field is effectively squeezed only if the control qubit is in the state $\vert 1\rangle$. On the other hand, if the state is $\vert 0\rangle$, ordinary harmonic evolution with frequency $\bar\omega_0$ takes place and when compensated as described below, turns this into a controlled-squeeze gate.

The setup we envision is described in Fig.~\ref{lomb.chap:setup}: the transmon qubit on the left is capacitively coupled with a $\lambda/4$ resonator terminated by a flux-tunable SQUID. This last component has been used to demonstrate the dynamical Casimir effect~\cite{lomb.svensson_period_2018,lomb.wilson_photon_2010}, and has been the subject of thorough investigation~\cite{lomb.dce_nori_pra,lomb.shumeiko_2013}.
The setup in Fig.~\ref{lomb.chap:setup} can be modelled with the Hamiltonian
\begin{equation}
\label{lomb.eq:1}
    \hat H (t) =  \frac{\omega_q}{2}\hat{\sigma}_z + \omega \hat a^{\dagger} \hat a  + \chi \hat a^{\dagger}\hat a \hat \sigma_{z} + g_d  \epsilon \sin(\omega_d t-\theta) (\hat a^{\dagger} +\hat a)^2  ,
\end{equation}
where $\hat a$ is the bosonic annihilation operator of the resonator mode and $\hat\sigma_z=|0\rangle\langle0|-|1\rangle\langle1|$ is the Pauli operator associated with the qubit. Here, $\omega_q$ is the frequency of the qubit, and $\chi$ is the dispersive coupling constant between the resonator and the transmon. The dependence of these constants on physically relevant parameters such as the Josephson energy, the capacitance of the transmon qubit, its charge energy, the coupling capacitance, the inductance and capacitance per unit length of the resonator and the parameters characterizing the external pumping.
The Hamiltonian in~(\ref{lomb.eq:1}) does not include non-linearities, which have been discussed in~\cite{lomb.ctrlsqz}, can be neglected when the ratio between the Josephson energy of the SQUID in the right and the inductive energy of the resonator ($E_{L,{\rm res}}/E_J$) is small.
In what follows we will work under this assumption and explain the effect of the Hamiltonian~(\ref{lomb.eq:1}), presenting later some results with experimentally realizable parameters discussing also the effect of losses, decoherence, and non-linearities. A similar Hamiltonian appears in the context of trapped ions where a gate that squeezes the motional degree of freedom of an ion depending on its internal state was proposed by modulating the amplitude of an optical lattice~\cite{lomb.paz2020}.

By looking at the Hamiltonian in Eq.~(\ref{lomb.eq:1}) one can see that it describes a harmonic oscillator with a resonance frequency $ \omega_{0,1}(t)$
that both varies in time and is conditioned on the qubit state ($\bar\omega_0=\omega+\chi$ when the qubit is in the state $|0\rangle$ or $\bar\omega_1=\omega-\chi$ when the qubit is in the state $|1\rangle$).
If the system is driven with $\omega_d = 2 \bar\omega_1$, and the qubit state is $|1\rangle$, then the parametric resonance is excited and the state of the resonator is squeezed. On the other hand, if the state of the qubit is $|0\rangle$, the resonator simply acquires a renormalized frequency due to the effect of the AC-Stark shift. Thus, in the frame rotating with frequency $\bar\omega_1$, the Hamiltonian~(\ref{lomb.eq:1}), within the rotating wave approximation (RWA), reads as
\begin{equation}
\hat H_{I} = \frac{1}{2}ig_d\epsilon(e^{-i\theta} \hat a^{2}-e^{i\theta}\hat a^{\dagger2})\otimes |1\rangle\langle1| + \tilde \Delta \hat a^\dagger \hat a \otimes |0\rangle\langle 0|,\label{lomb.HI}\end{equation}
where $\tilde \Delta \approx \Delta (1 - 1/2(g_d \epsilon/\Delta)^{2}) + {\cal O}((g_d \epsilon/\Delta)^{4})$.

The temporal evolution operator associated with the above Hamiltonian is
\begin{equation}\exp(-i\hat H_{I} t) = \hat S(r,\theta)\otimes |1\rangle\langle1|+ \hat U_0(\varphi) \otimes |0\rangle\langle0| \equiv  \hat U(r, \theta,\varphi),\label{lomb.evol}\end{equation}
where $\hat{S}(r,\theta)$ is the squeezing operator defined above, with the squeezing parameter $r=-g_d\epsilon t$ and the squeezing angle $\theta$ set by the phase of the driving. Above, $\hat U_0(\varphi)$ is the evolution operator of an oscillator with frequency $\tilde \Delta$, which during a time $t$, induces a rotation in phase space in an angle $\varphi = \tilde\Delta t$. For the above evolution operator $\hat U(r,\theta,\varphi)$ in Eq.~(\ref{lomb.evol}) to be a true controlled-squeeze gate, it is necessary to compensate for the free evolution $\hat U_0(\varphi)$. This can be done, at least in two different ways. First, one can  turn off the magnetic driving and then wait a time $\tau$ chosen in such a way that $\Delta \tau + \varphi = 2 k \pi$, for some integer $k$. After this, as $\hat U_0(\Delta \tau) \hat U_0(\varphi) = 1$, the evolution operator is the desired controlled-squeeze gate: $\textbf{C-Sqz}(r,\theta) = \hat U_0(\Delta \tau) \otimes |0\rangle \langle 0| \otimes \hat U(r,\theta,\varphi)$. A different alternative, that's not required to turn off the magnetic driving, is to use the non-compensated controlled-squeeze gate $\hat U(r,\theta,\varphi)$ and to choose the subsequent operations to depend on the rotation angle $\varphi$. We will follow this second strategy below.

Now we show how to use the above result in order to encode an arbitrary qubit state in a resonator making the errors induced by photon losses detectable~\cite{lomb.ofek2016,lomb.ni2023beating,lomb.Grimsmo2020,lomb.teoh2023dual}. We choose the encoding in such a way that the logical states $|0\rangle$ and $|1\rangle$  are represented by the states  $|\chi_+\rangle$ and $|\chi_-\rangle$ of the resonator built as even and odd superpositions of states squeezed along two orthogonal directions in quadrature space, i. e.,
\begin{align}
    |\chi_{\pm}\rangle=\frac{1}{\sqrt{2}c_\pm}(|r,\tilde{\theta}\rangle \pm |r,\tilde\theta+\pi\rangle)\label{lomb.chipm},
\end{align}
where $|r,\tilde\theta\rangle$ is a one-mode squeezed state and the constant $c_\pm=\sqrt{1\pm1/\sqrt{\cosh 2r}}$.
From this it is simple to see that the states $|\chi_+\rangle$ and $|\chi_-\rangle$ have similar properties to the four-legged cat~\cite{lomb.ofek2016} states as they are respectively superpositions of $4n$ and $4n+2$ photon states, which implies that when losing a photon the encoded state still stores a coherent superposition and the error can be detected by a subsequent parity measurement of the photon number inside the resonator.

To prepare a general encoded state we should start with an arbitrary qubit state $|\Psi_Q\rangle=\alpha|0\rangle+\beta|1\rangle$ and the resonator in the vacuum. Then we apply the following sequence: i) Apply a Hadamard gate to the qubit (transforming $|0\rangle\to(|0\rangle+|1\rangle)/\sqrt{2}$ and $|1\rangle\to(|0\rangle-|1\rangle)/\sqrt{2}$); ii) Apply the non-compensated controlled-squeeze gate $\hat U(r,\theta,\varphi)$ defined in Eq.~(\ref{lomb.evol}); iii) Apply a $\pi$-rotation to the qubit;  iv) Apply the operator $\hat U(r,\theta + 2 \varphi + \pi, \varphi)$; v) Apply a $\pi$-rotation to the qubit; and vi) Apply another Hadamard gate to the qubit. After this sequence the combined qubit-resonator state will be
\begin{equation}
    |\Psi_{QR}\rangle =\frac{1}{\sqrt{2}}|0\rangle\left(\alpha\ {c}_{+}|\chi_{+}\rangle+\beta\ {c}_{-}|\chi_{-}\rangle\right)
    +\frac{1}{\sqrt{2}}|1\rangle\left(\alpha\ {c}_{-}|\chi_{-}\rangle+\beta\ {c}_{+}|\chi_{+}\rangle\right)\label{lomb.psiQR}
\end{equation} where $|\chi_{+}\rangle$ and $|\chi_{-}\rangle$ are the above defined ones with $\tilde \theta = \theta + 2 \varphi$.
If we measure $\hat{\sigma}_z$ for the qubit, we obtain the results $\pm1$, which respectively identify the states $|0\rangle$ or $|1\rangle$, with probability $P_\pm= 1/2 (1\pm P_z/\sqrt{\cosh(2r)})$, where $P_z= \alpha^2-\beta^2$ is the z-component of the polarization vector of the qubit which identifies its state in the Bloch sphere. For each result, the state of the resonator turns out to be
$|\Psi_{R}^\pm\rangle=(\alpha{c}_{\pm}|\chi_{\pm}\rangle+\beta{c}_{\mp}|\chi_{\mp}\rangle)/\sqrt{|\alpha|^2 c_\pm^2  +  |\beta|^2 c_\mp^2}$.
For each result $\sigma_z = \pm 1$ the resonator stores the encoded states $|\Psi_R^\pm\rangle$ whose fidelity with respect to ideal state $|\tilde{\Psi}_Q^\pm\rangle = \alpha| \chi_\pm\rangle + \beta|\chi_\mp\rangle$ is $F_\pm = |\langle\tilde{\Psi}_Q^\pm|\Psi_R^\pm\rangle|^2$. The average fidelity for the complete encoding protocol is $\bar F = P_+ F_+ + P_- F_-$ which can be expressed as

\begin{equation} \bar F = \frac{1}{2}(1 + P_z^2) + \frac{1}{2}(1 - P_z^2) \sqrt{1 - \frac{1}{\cosh(2r)}}. \label{lomb.averageF}\end{equation}
The lowest fidelity states are those on the equator, i.e. when $P_z= 0$, and, in the limit of high squeezing, we find that $\bar F\sim 1 - e^{-2r}(1-P_z^2)/4$. Clearly, to enforce a high fidelity for every state we need the squeezing factor to be large enough. In fact, $\bar F\geq 0.995$ requires $r\geq2$.

We analysed the implementation of the above encoding protocol considering material properties that have been achieved in systems similar to the one proposed~\cite{lomb.delsing_new,lomb.ganjam2024surpassing}. For this, we chose the resonator frequency $\omega/(2\pi)=6\text{GHz}$, the qubit frequency $\omega_q/(2\pi)=4\text{GHz}$, the driving coupling $g_d=50\text{MHz}$, the driving amplitude  $\epsilon=0.15$, and the qubit coupling strength $\chi/(2\pi)=8\text{MHz}$. The controlled-squeeze gate is applied during $200\rm  n\text{s}$, resulting in a squeezing $r\sim 1.5$.

We included the effect of losses and decoherence modeled through a master equation describing thermal contact between the qubit-resonator system and a bath at $60\text{mK}$. For the relaxation time-scales, we choose  values which lie in between the ones reported in Ref. \cite{lomb.delsing_new} and the most recent one~\cite{lomb.ganjam2024surpassing}: a qubit relaxation time-scale $\tau_q=200\mu\text{s}$, a resonator damping time $\tau_r=200\mu\text{s}$, and a qubit dephasing time-scale of $\tau_\phi=10\mu\text{s}$.

Including non-linearities, losses, and decoherence, we studied the average fidelity and purity decay for arbitrary states, where the dependence of these quantities with the azimuthal angle $\vartheta$ is calculated for two typical values of the
polar $\phi$-angle in the Bloch sphere. For the above parameters we find average fidelity between  $96.2\%$ (for the states in the equator) and $98.9\%$ (for states in both poles of the Bloch sphere). For the ideal average fidelity given by Eq.~(\ref{lomb.averageF}) where dissipation and dephasing are not considered,  the obtained values are $1\%$ higher than those arising from numerical simulations that include losses, decoherence, and non-linearities.
 Purity ranges between  $97.3\%$ (in the equator) and $99.3\%$ (in the poles).

The imperfection in the compensation angle introduces a systematic error in the gate that is very small. Thus, the maximum achievable fidelity in the absence of losses and decoherence in the equator for the squeezing parameter $r = 1.5$ is $97\%$ (while it would be $97.5\%$ for the perfect unitary controlled-squeeze gate).

\section{Summary}
\label{lomb.chap:summary}

We have studied some of the consequences of the dynamic Casimir effect on thermodynamics and quantum information in a variety of systems. The goal has been, on one hand, to identify in which processes this effect can negatively impact and, in such cases, whether it can be mitigated; and on the other hand, to determine which processes can benefit from its application and how to maximize it. Among these processes, we have considered the energy transfer from phonons in a moving wall to photons in a cavity, where the DCE is exploited in a virtuous way~\cite{lomb.nos_conversion}; we have also examined the case of a thermal machine based on branches, where the DCE appears as a byproduct that negatively impacts performance~\cite{lomb.Otto_nos}. From the perspective of quantum information, we  have additionally studied the entanglement between two cavities in relative oscillatory motion, where the DCE negatively affects information storage~\cite{lomb.nos_entangled}. However, we have further observed that the effect can be harnessed in superconducting circuits to create gates and quantum states that facilitate the processing and storage of quantum information~\cite{lomb.ctrlsqz}.

We have studied the consequences of the DCE in an optomechanical cavity undergoing an Otto cycle. To this end, we have found an analytical expression for the energy generated by the DCE as a function of the wall's trajectory. We have observed that this energy is always positive, regardless of the direction of motion, and is proportional to the square of the velocity, highlighting its quantum friction character. Furthermore, we have shown that this leads to a significant decrease in the cycle's efficiency for very short operation times. As a result, the power delivered by an engine using this cycle increases as the operation time decreases, but it reaches a maximum. Beyond this point, the frictional energy due to the DCE becomes dominant, and the work delivered decreases abruptly, ultimately reaching zero at a finite operation time.

To conclude, we have demonstrated how the squeezing generated by the DCE can be exploited to produce a controlled-squeeze gate. Furthermore, we have presented a concrete experimental proposal based on a superconducting waveguide terminated by a SQUID and controlled by a capacitively coupled transmon qubit. We have shown that this implementation allows for the generation of unitary evolution of quantum-controlled squeezing by the qubit, and we have analysed its experimental feasibility through simulations with realistic parameters. Additionally, we have studied the use of this gate for generating squeezed superposition states with significant applications in metrology. Finally, we have shown that the controlled-squeeze gate, in conjunction with a displacement operation, can generate any other unitary operation on the cavity, meaning it is universal, making it a valuable tool for quantum computing.

\section*{Acknowledgment}

This work was funded by Agencia Nacional de
Promocion Científica y Tecnológica (ANPCyT), Consejo
Nacional de Investigaciones Cientıficas y Técnicas
(CONICET), and Universidad de Buenos Aires
(UBA). We are deeply grateful to N. Del Grosso, J.P. Louys Sanso, F.D Mazzitelli, C. Velasco, R. Corti\~nas, and Juan Pablo Paz, with whom we have developed the research reviewed in this chapter.

\renewcommand{\bibname}{References}
\begingroup
\let\cleardoublepage\relax

\endgroup

\end{document}